\documentclass[fleqn,usenatbib]{mnras}

\usepackage{newtxtext,newtxmath}

\usepackage[T1]{fontenc}
\usepackage{ae,aecompl}

\usepackage{graphicx}	
\usepackage{amsmath}	
\usepackage{amssymb}	
\usepackage{multirow}

\newcommand{\aref}[1]{\hyperref[#1]{Appendix~\ref{#1}}}

\newcommand{\jcompphys}{J.~Comp.~Phys}

\newcommand{\vecrho}{\boldsymbol{\rho}}
\newcommand{\matA}{\boldsymbol{\mathsf{M}}}

\usepackage{scalerel,tikz}
\usetikzlibrary{svg.path}
\definecolor{orcidlogocol}{HTML}{A6CE39}
\tikzset{orcidlogo/.pic={
 \fill[orcidlogocol] svg{M256,128c0,70.7-57.3,128-128,128C57.3,256,0,198.7,0,128C0,57.3,57.3,0,128,0C198.7,0,256,57.3,256,128z};
 \fill[white] svg{M86.3,186.2H70.9V79.1h15.4v48.4V186.2z}
 svg{M108.9,79.1h41.6c39.6,0,57,28.3,57,53.6c0,27.5-21.5,53.6-56.8,53.6h-41.8V79.1z M124.3,172.4h24.5c34.9,0,42.9-26.5,42.9-39.7c0-21.5-13.7-39.7-43.7-39.7h-23.7V172.4z}
 svg{M88.7,56.8c0,5.5-4.5,10.1-10.1,10.1c-5.6,0-10.1-4.6-10.1-10.1c0-5.6,4.5-10.1,10.1-10.1C84.2,46.7,88.7,51.3,88.7,56.8z};
}}
\newcommand\orcidicon[1]{\href{https://orcid.org/#1}{\mbox{\scalerel*{
\begin{tikzpicture}[yscale=-1,transform shape]
\pic{orcidlogo};
\end{tikzpicture}
}{|}}}}

\title[Transitional disc dust dynamics]{Dynamics of small grains in transitional discs}

\author[Krumholz, Ireland, \& Kratter]{Mark R. Krumholz$^{\orcidicon{0000-0003-3893-854X}\,1,2,3,4}$\thanks{E-mail: mark.krumholz@anu.edu.au},
Michael J. Ireland$^{1}$\thanks{E-mail: michael.ireland@anu.edu.au}, and
Kaitlin M. Kratter$^{5}$
\\
$^{1}$Research School of Astronomy and Astrophysics, Australian National University, Canberra, ACT 2611, Australia\\
$^{2}$ARC Centre of Excellence in All-Sky Astrophysics (ASTRO-3D), Canberra, ACT 2611, Australia\\
$^{3}$Universit\"at Heidelberg, Zentrum f\"ur Astronomie, Institut f\"ur Theoretische Astrophysik, 69120 Heidelberg, Germany\\
$^{4}$Max Planck Institute for Astronomy, K\"onigstuhl 17, 69117 Heidelberg, Germany\\
$^{5}$Department of Astronomy and Steward Observatory, University of Arizona, Tucson, Arizona 85721
}

\date{Accepted XXX. Received YYY; in original form ZZZ}

\pubyear{2019}

\begin{document}
\label{firstpage}
\pagerange{\pageref{firstpage}--\pageref{lastpage}}
\maketitle

\begin{abstract}
Transitional discs have central regions characterised by significant depletion of both dust and gas compared to younger, optically-thick discs. However, gas and dust are not depleted by equal amounts: gas surface densities are typically reduced by factors of $\sim 100$, but small dust grains are sometimes depleted by far larger factors, to the point of being undetectable. While this extreme dust depletion is often attributed to planet formation, in this paper we show that another physical mechanism is possible: expulsion of grains from the disc by radiation pressure. We explore this mechanism using 2D simulations of dust dynamics, simultaneously solving the equation of radiative transfer with the evolution equations for dust diffusion and advection under the combined effects of stellar radiation and hydrodynamic interaction with a turbulent, accreting background gas disc. We show that, in transition discs that are depleted in both gas and dust fraction by factors of $\sim 100-1000$ compared to minimum mass Solar nebular values, and where the ratio of accretion rate to stellar luminosity is low ($\dot{M}/L \lesssim 10^{-10}$ $M_\odot$ yr$^{-1}$ $L_\odot^{-1}$), radiative clearing of any remaining $\sim 0.5$ $\mu$m and larger grains is both rapid and inevitable. The process is size-dependent, with smaller grains removed fastest and larger ones persisting for longer times. Our proposed mechanism thus naturally explains the extreme depletion of small grains commonly-found in transition discs. We further suggest that the dependence of this mechanism on grain size and optical properties may explain some of the unusual grain properties recently discovered in a number of transition discs. The simulation code we develop is freely available.
\end{abstract}




\section{Introduction}

Transitional discs are so-named as the stage in evolution of a single star's protoplanetary disc in-between the optically thick Class~II and the optically thin Class~III stage, within the framework of inside-out disc clearing \citep{Alexander14}.
Although sometimes confused with circumbinary discs \citep{Espaillat07, Ireland08}, these discs remain an important stage in single star evolution \citep[e.g.][]{RuizRodriguez16} 
and may be a signpost for giant planet or multiple giant planet formation \citep{Zhu:2011,DodsonRobinson11}. Even if they are not a sign of planet formation in all systems, the phase when the disc is transitioning from optically thick to thin in the giant planet formation region is generally when giant planets are at their most detectable from high contrast surveys, as the recent example of 
PDS~70 has shown \citep{keppler_discovery_2018,wagner_magellan_2018}.

However, many disc features are ambiguous, and the planetary nature of such features remains highly debated in numerous discs including
T~Cha \citep{huelamo_companion_2011,cheetham_near-ir_2015}, 
the infrared emission in LkCa~15 \citep{kraus_lkca_2012,thalmann_resolving_2016}, HD~100546 \citep{quanz_young_2013,rameau_optical/near-infrared_2017} and HD~169142 \citep{biller_enigmatic_2014,reggiani_discovery_2014,ligi_investigation_2018}. Much of the confusion surrounding these features is attributable to the inadequacy of simple dust models to explain the observed scattered light emission. 
T~Cha had very bright emission from forward scattering, LkCa~15 had both bright and red emission appearing in a forward scattering geometry \citep{ireland_lkca_2014,thalmann_lkca_2015,currie_lkca_2019}, and HD~169142 required emission from extremely small or quantum heated grains in a disc where 
micron sized grains were largely absent \citep{Birchall19}. These observed complex grain distributions motivates this paper, which considers grain segregation processes in transitional discs.

It is imperative to place the systems we model in the broader context of the entire class of transition discs. Even when classified exclusively via their SEDs (the original definition), they comprise a heterogeneous set of sources \citep{Espaillat14,Ercolano17a}. They contain inner cavities of a range of sizes that do not show up uniformly in multi-wavelength observations (e.g. millimeter, or IR polarized emission). Disk masses inferred from dust continuum observations also vary greatly. Finally, they exhibit a wide range of accretion rates, from upper limits of order $\dot{M} \lesssim 10^{-11} M_{\odot}/\rm{yr}^{-1}$ up to detections of accretion rates similar to those of Class I-II disks, $\dot{M} = 10^{-7} M_{\odot}/\rm{yr}^{-1}$ \citep{Najita15}. It is thus likely that the SED-identified sample of transitions discs is not described by a single evolutionary channel. Our use of the term ``transition disc" is not meant to represent the entire class, but is a stand-in for a subset of these objects that are undergoing the final stages of disc dispersal, with low gas and dust masses, and small accretion rates.

Most previous work on grain segregation in gaseous discs considers the effects of either settling or gas pressure gradients combined with variable dust stopping times. For example, \cite{Takeuchi:2002} considered the radial flow of dust particles in a disc where radiation pressure was neglected, and the disc evolved viscously with a constant $\alpha$ prescription. In transition discs, the lower gas densities mean that significantly smaller grains settle to the midplane, and as the disc becomes optically thin, radiation pressure on grains can become important. \cite{Takeuchi:2001} considered the joint effects of a gas pressure gradient and radiation pressure in a disc that is optically thin throughout, neglecting the effects of gas accretion. They found that radiation could remove $<$100\,$\mu$m dust grains from the inner 10s of AU of a 10\,M$_{\earth}$ disc on $\sim$kyr timescales, and segregate grains according to their size. 
\cite{takeuchi_surface_2003} modeled the combined effects of radiation pressure and accretion, focusing on discs similar to a minimum mass solar nebula, where surface outward dust motion was almost negligible compared to the motion of the bulk of the disc inwards with the gas accretion. 
\cite{tazaki_outward_2015} considered the motion of grains with a large radiative cross section in the surface layers of a minimum mass solar nebula (10\,M$_{\rm J}$ disc). They found that compact grains were not efficiently transported by radiation pressure, while smaller grains could be. \citet{Kenyon16a} considered a variety of mechanisms to remove dust from transition discs, and concluded either that planet formation must leave behind far fewer small grains than one might naively expect, or that some other mechanism, for example drag-induced accretion, must be available to clear dust from discs. \citet{Owen19a} suggest that radiation pressure operating on dust grains that reach the edge of a photoevaporation-driven cavity might efficiently remove them.

In this paper, we extend previous work by jointly considering the effects of gas pressure gradients and gas flows, stopping times that depend on grain radius, radiation and turbulent diffusion in order to study the dust temporal evolution during the transitional disc period of a protoplanetary disc.

\section{Dust evolution model}
\label{sec:dust_evol}

We are interested in modeling the evolution of a population of dust grains orbiting a central star of mass $M$ and luminosity $L$ in a background gas disc. We will treat grains as simple spheres of radius $a$ and density $\rho_s\approx 1$ g cm$^{-3}$, and we ignore coagulation and shattering, so that the total mass of dust grains of any given size is conserved. The grains move in response to radiation forces and to drag forces exerted by the gas; below we will assume that the stopping time of grains is small, so that these forces are always in balance and the grains are at their terminal velocity. We further separate grain motion into two types: systematic drift at a terminal velocity determined by force balance considering only the bulk velocity of the gas, and random motions as a result of local drag forces due to turbulent motions in the gas, which we approximate as a diffusion process. Under these assumptions, the density $\rho_{d,a}$ of dust grains of radius $a$ evolves following an advection-diffusion equation,
\begin{equation}
\label{eq:advec_diff}
\frac{\partial \rho_{d,a}}{\partial t} + \nabla \cdot \left(\rho_{d,a} \mathbf{v}_{d,a} + \mathbf{j}_{d,a} \right) = 0.
\end{equation}
Here $\mathbf{v}_{d,a}$ is the bulk velocity and $\mathbf{j}_{d,a}$ is the flux due to turbulent diffusion. This formulation of the evolution equations is identical to that proposed by \citet{Takeuchi:2002}.

We now proceed to calculate the bulk velocity and diffusion coefficient. In the following discussion, we use $\varpi$ and $z$ to denote the radial and vertical position in a cylindrical coordinate system centred on the star, and $r = \sqrt{\varpi^2+z^2}$ and $\theta = \sin^{-1}\left(z/r\right)$ to denote the corresponding radius and angle in a spherical coordinate system.  We assume that the system is symmetric in the azimuthal angle. In the equations that follow, we aim to differentiate between radial and cylindrical components; a novel aspect of our model is the full treatment of two dimensional drift in the presence of radiation pressure.

\subsection{Gas disc model}
\label{ssec:gas_disc}

As a first step, we specify our model for the gas disc through which the grains flow. We treat the gas disc as constant in time. The run of surface density and temperature through the disc are described by powerlaws,
\begin{eqnarray}
\Sigma_g(\varpi) & = & \epsilon \Sigma_{\rm MMSN} \left(\frac{\varpi}{\mathrm{AU}}\right)^p \\
T(\varpi) & = & T_0 \left(\frac{\varpi}{\mathrm{AU}}\right)^q,
\label{eq:Tvsr}
\end{eqnarray}
where $\epsilon$ is a dimensionless factor that scales the mass in the disc to that of the minimum mass Solar nebula (MMSN), $\Sigma_{\rm MMSN} \approx 2200$ g cm$^{-2}$, $T_0$ is the disc temperature at 1 AU ($T\approx 120$ K for a Sun-like star), and $p$ and $q$ are constant. We are interested in transition discs, for which $\epsilon \ll 1$. Standard values for us are $p=-3/2$ and $q=-3/7$, as expected for a \cite{Chiang:1997} passive disc profile, but we will also consider the case of discs that have been depleted at small radii, and thus have $p=0$. Our temperature profile neglects the effects of dust settling, and we further simplify the situation by assuming that the gas is vertically isothermal. Under this assumption the sound speed $c_s$ is constant with $\varpi$, and the scale height $h_g$ follows the usual relation $h_g = c_s/\Omega_{\rm K,mid}$, where 
\begin{equation}
\Omega_{\rm K,mid} = \left(\frac{G M}{\varpi^3}\right)^{1/2}
\end{equation}
is the Keplerian angular velocity at the midplane.

Assuming the gas disc is in hydrostatic equilibrium, we have
\begin{eqnarray}
0 & = & -\frac{GMz}{r^3} - \frac{1}{\rho_g}\frac{\partial P_g}{\partial z} \\
\rho_g &=& \frac{\Sigma_g}{h_g} \exp\left(-\frac{z^2}{2h_g^2}\right)\nonumber \\
&=&2.7\times 10^{-9}\epsilon \left(\frac{\varpi}{{\rm AU}}\right)^{-39/14} \exp\left(-\frac{z^2}{2h_g^2}\right)\mbox{ g cm}^{-3}
\label{eq:rhog_fiducial}
\end{eqnarray}
where $P_g = \rho_g k_B T/\mu m_{\rm H}$ is the gas pressure under the reasonable approximation of an ideal gas, $\mu$ is the mean molecular weight in units of the hydrogen mass $m_{\rm H}$, $\rho_g$ is the gas volume density, and the final equation contains fiducial MMSN values and scalings \citep{ChiangYoudrev09}. Also assuming hydrostatic equilibrium in the radial direction, we have \citep{ChiangYoudrev09}
\begin{equation}
\varpi\Omega_g^2 - \frac{GM\varpi}{r^3} - \frac{1}{\rho_g}\frac{\partial P_g}{\partial \varpi}=0,
\end{equation}
where $\Omega_g$ is the gas angular velocity. Expanding this equation in powers of $z/r$ and keeping terms up to order $(z/r)^2$ \citep{Takeuchi:2002}, we have
\begin{equation}
\label{eq:omega_g}
\Omega_g = \Omega_{\rm K,mid} \left[1+\frac{1}{2}\left(\frac{h_g}{r}\right)^2\left(p-(q+3)/2+q+\frac{q}{2}\frac{z^2}{h_g^2}\right)\right] 
\end{equation}
This is conveniently expressed as:
\begin{eqnarray}
\Omega_{\rm K} & \approx & \Omega_{\rm K,mid} \left(1-\frac{3}{4}\frac{z^2}{r^2}\right)\\
\Omega_g &= &\Omega_{\rm K} (1-\eta)^{1/2}\\
\eta  &=& -\left(\frac{h_g}{r}\right)^2\left(p-\frac{q+3}{2}+q+\frac{q+3}{2}\frac{z^2}{h_g^2}\right).
\end{eqnarray}
Here $\Omega_{\rm K}$ is the Keplerian speed a height $z$ above the midplane and $\Omega_g$ is the gas angular velocity, which is smaller than the Keplerian velocity by a factor of $\sqrt{1-\eta}$ due to the effects of pressure support.

In order to solve for the dust velocity we will require an expression for the gas radial velocity. Specifically we require a height-dependent velocity to be self-consistent. For simplicity we choose 
a simple model consisting of an $\alpha$-disc plus an optional constant inflow rate for this purpose. We begin with a pure $\alpha$-disc case, starting with the azimuthal component of the momentum equation:
\begin{eqnarray}
\lefteqn{2\pi \rho_g \left(v_{g,\varpi} \frac{\partial}{\partial \varpi}+v_{g,z}\frac{\partial}{\partial z}\right) \left(\varpi^2\Omega_g\right)
}
\\ \nonumber
& = &
2\pi \left[\frac{\partial}{\partial \varpi}\left(\varpi^2\rho_g\nu\frac{\partial\Omega_g}{\partial\varpi}\right)+\frac{\partial}{\partial z}\left(\varpi^2\rho_g\nu\frac{\partial\Omega_g}{\partial z}\right)\right],
\end{eqnarray}
where $\nu$ is the kinematic viscosity. The term $v_{g,z}(\partial/\partial z) (\varpi^2\Omega_g)$ is smaller than $v_{g,\varpi}$ by $h_g/\varpi$, and so can be neglected \citep{Takeuchi:2002}. Solving for radial gas velocity, and retaining terms up to order $(h_g/r)^2$, we find
\begin{equation}
\label{eq:vgvarpi_turb}
v_{g,\varpi,\rm turb} = -\alpha \frac{h_g}{r} c_s \left[3\left(p-\frac{q+3}{2}\right)+2q+6+\frac{5q+9}{2}\left(\frac{z}{h_g}\right)^2\right],
\end{equation}
where $\alpha = \nu/(c_s h_g)$ is the usual dimensionless viscosity
(see also \citealt{Keller:2004}), and the subscript ``turb'' indicates that this is the radial velocity associated with the turbulent flow. \autoref{eq:vgvarpi_turb} gives the gas radial velocity as a function of cylindrical radius $\varpi$ (implicitly, through $r$ and $z$) and height $z$ above the midplane.

The dominant driver of accretion in transitional disks remains uncertain. The exposed inner rims and low surface density regions might  be sufficiently ionized via cosmic rays or stellar photons to sustain magnetorotational instability -driven turbulence (hereafter MRI) \citep{ChiangMC2007,Mohanty:2018}. Alternatively, other mechanisms such as wind-driven accretion may dominate \citep[e.g.,][]{Bai13,Turner14}. These tend to produce laminar rather than turbulent flows, which we add to our model simply by adding a laminar inflow velocity $v_{g,\varpi,\rm lam}$ to the turbulent one. This laminar velocity depends only on $\varpi$ (i.e., it is constant with height within the disc), and is parameterised by the accretion rate $\dot{M}$, which we take to be constant with radius. Thus our net model for the gas velocity is
\begin{equation}
    \label{eq:vgvarpi}
    v_{g,\varpi} = v_{g,\varpi,\rm turb} - \frac{\dot{M}}{2\pi \varpi \Sigma_g},
\end{equation}
where $v_{g,\varpi,\rm turb}$ is given by \autoref{eq:vgvarpi_turb}, and the dimensionless viscosity $\alpha$ and mass accretion rate $\dot{M}$ are left as free parameters that we will vary below. While simplified (either the MRI or disk-winds might well produce height dependent inflow rates), we expect our model to nevertheless capture the overall trends associated with an extra source of radial drift for the gas and dust.

\subsection{Dust velocity}

Now consider a dust grain, working in a reference frame co-rotating with the grain at azimuthal velocity $v_{d,\phi}$. In this frame, the equation of motion in the $(\varpi,z)$ plane, considering only the bulk velocity of the gas and not its small-scale turbulent motion, and neglecting Poynting-Robertson drag, is
\begin{equation}
\frac{d}{dt}\mathbf{v}_{d,a} = \frac{v_{d,\phi}^2}{\varpi} \hat{\varpi} + \left(\beta-1\right)\frac{GM}{r^2} \hat{r} + \frac{3 \mathbf{F}_{\rm drag}}{4\pi a^3 \rho_s},
\end{equation}
where $\beta$ is the ratio of outward radiation pressure force to inward gravitational force and $\mathbf{F}_{\rm drag}$ is the force exerted by gas drag. The first term here represents the centrifugal force, the second is the radiative minus gravitational force, and the third is the gas drag force. We omit the Coriolis force because it exerts forces only in the azimuthal direction.

\subsubsection{Radiation force}

In general the radiation pressure force must be determined by integrating in frequency. Following \citet{Wolfire87a}, if the central star has specific luminosity $L_\nu$ and we neglect the scattered and dust-reprocessed component of the radiation field compared to the direct stellar field, then the radiation pressure force on a grain is
\begin{equation}
\label{eq:frad}
F_{\rm rad} = \int \frac{L_{\nu}}{c} e^{-\tau_\nu} \frac{a^2}{4 r^2} \left[Q_{a,\nu}^A + \left(1 - g_{a,\nu}\right) Q_{a,\nu}^S\right] \, d\nu,
\end{equation}
where $Q_{a,\nu}^A$ and $Q_{a,\nu}^S$ are the absorption and scattering efficiencies for grains of size $a$ at frequency $\nu$, and $g_{a,\nu}$ is the cosine of the mean scattering angle (with $g_{a,\nu} = 1$ indicating complete forward scattering). The optical depth from the stellar surface at radius $r_*$ to the radial distance $r$ of the grain is
\begin{equation}
\tau_\nu = \int_{r_*}^r \int_0^{\infty} \frac{3}{4 a} \frac{\rho_{d,a}}{\rho_s} \left(Q_{a,\nu}^A + Q_{a,\nu}^S\right) \, da \, dr',
\end{equation}
where $\rho_s$ is the density of the grains. 
Evaluation of \autoref{eq:frad} in general must be done numerically, and is numerically expensive if one requires high frequency resolution. 
However, the simplest application and one with broad applicability is for grains of size $a$ much larger than the wavelength of photons at the peak of the stellar spectral energy distribution. Specifically, if the stellar effective temperature is $T_*$, yielding a wavelength of peak emission per unit wavelength $\lambda_* \approx hc/(4.965 k_B T_*)$, grains will be in the limit of geometric optics, $Q_{s,\nu}^A + (1-g_{a,\nu}) Q_{s,\nu}^S= 1$, if their size obeys
\begin{equation}
a \gg \frac{\lambda_*}{2\pi} \approx 0.077 \left(\frac{T_*}{6000\mbox{ K}}\right)\,\mu\mathrm{m}.
\end{equation}
For such grains, the radiation force and optical depth reduce to
\begin{eqnarray}
F_{\rm rad} & = & \frac{L}{c} e^{-\tau} \frac{a^2}{4r^2} \\
\tau & = & \int_0^\infty \frac{3}{4a} \int_{r_*}^r \frac{\rho_{d,a}}{\rho_s} \, dr' \, da \equiv \tau.
\label{eq:tau_geom}
\end{eqnarray}
With this simplification, it is convenient to express the ratio of radiation pressure force to gravitational force as simply
\begin{equation}
\label{eq:beta_geom}
\beta = \beta_{a,0} e^{-\tau},
\end{equation}
where
\begin{equation}
\label{eq:betazero}
\beta_{a,0} = \frac{3 L}{16 \pi G M c \rho_s a}
= 0.57 \left(\frac{L/M}{L_\odot/M_\odot}\right) \left(\frac{\rho_s}{1\;\mathrm{g}\;\mathrm{cm}^{-3}}\right)^{-1} \left(\frac{a}{1\;\mu\mathrm{m}}\right)^{-1}
\end{equation}
is the ratio of radiative to gravitational force for a grain of size $a$ exposed to the full, unshielded luminosity of the star. 

\subsubsection{Drag force}

We compute the drag force under the assumption that the grains are small enough to obey the Epstein drag law,
\begin{equation}
\mathbf{F}_{\rm drag} = -\frac{4}{3}\pi \rho_g a^2 c_s \Delta\mathbf{v},
\end{equation}
where $\rho_g$ is the gas density, $c_s$ is the gas sound speed, and $\Delta\mathbf{v}$ is the relative velocity of the gas and dust. Combining the radiative and drag terms, the total equation of motion for a single grain of size $s$ is
\begin{equation}
\label{eq:grain_eom}
\frac{d}{dt} \mathbf{v}_{d,s} = \frac{v_{d,\phi}^2}{\varpi} \hat{\varpi} + \left(\beta-1\right) \frac{G M}{r^2} \hat{r} - \frac{\Delta\mathbf{v}}{t_s},
\end{equation}

where
\begin{equation}
t_s = \frac{\rho_s a}{\rho_g c_s}
\end{equation}
is the usual stopping time. Below it will be more convenient to work with the dimensionless stopping time
\begin{equation}
T_s = t_s \Omega_{\rm K}.
\label{eq:stopping_time}
\end{equation}
Note that, although we will not write this out explicitly for reasons of compactness, it is important to recall that $T_s$ is a function of the grain size $a$.

If we limit ourselves to considering grains of size $a$ such that $T_s \ll 1$ near the midplane, then over timescales longer than an orbit the left hand side of \autoref{eq:grain_eom} approaches zero in the radial and vertical direction as the grains reach terminal velocity. The condition for this to hold is that
\begin{equation}
a \ll \frac{\rho_{g,\rm mid} c_s}{\rho_s \Omega_{\rm K,mid}} \approx 880 \epsilon \left(\frac{\varpi}{\mathrm{AU}}\right)^p\;\mathrm{cm},
\end{equation}
where $\rho_{g,\rm mid}$ is the midplane gas density, and in the second step we have taken $M=M_\odot$ and inserted our fiducial value for $\rho_g$ (\autoref{eq:rhog_fiducial}). Thus our approximations that grains can be treated in the geometric optics limit and that they reach terminal velocity quickly are valid over a wide range of grain sizes -- from $\approx 0.1$ $\mu$m up to cm to m, depending on the value of $\epsilon$.

In the vertical direction we find the dust terminal velocity to first order is 
\begin{eqnarray}
    \Delta v_z = t_s\left(\beta-1\right)v_{\rm K}^2\frac{z}{r}\\
    v_{d,z} = T_s\left(\beta-1\right)\Omega_{\rm K} z \label{eq:vz},
\end{eqnarray}
where we have taken $v_{g,z} = 0$ to arrive at the second equation.
To obtain an expression for the radial drift, first consider the azimuthal component of the dust momentum equation. Because the dominant source of angular momentum is simply Keplerian motion, we can relate the rate of angular momentum change to the drift rate of solids \citep{PinillaYoudin17}:
\begin{equation}
\frac{d(\varpi v_{d,\phi})}{dt} \simeq -\frac{d(\varpi v_K)}{d\varpi}v_{d,\varpi} =- \frac{1}{2}v_K v_{d,\varpi}
\end{equation}
We can use this relation to replace the L.H.S of the $\phi$ component of \autoref{eq:grain_eom}, which simplifies to
\begin{equation}
\label{eq:deltav_phi}
\left(v_{d,\phi}-v_{g,\phi}\right)= -\frac{T_s}{2}v_{d,\varpi}.
\end{equation}
We can use this relative velocity in the azimuthal direction to solve for the dust terminal velocity in the (cylindrical) radial direction. From \autoref{eq:grain_eom} we have:
\begin{equation}
    \Delta v_{\rm \varpi} = t_s\left[\frac{v_{d,\phi}^2}{\varpi}+\left(\beta-1\right)\Omega_
    {\rm K}^2{\varpi}\right]
    \label{eq:simple_eom_cyl}
\end{equation}
We require a linearized expression for the dust azimuthal velocity $v_{d,\phi}$. Following \citet{Takeuchi:2002} and \citet{PinillaYoudin17}, we can remove higher order terms by relating $v_{d,\phi}$ and $v_K$ 
\begin{eqnarray}
    v_{d,\phi}^2-v_K^2 &=& (v_{d,\phi} +v_K)(v_{d,\phi}-v_K) \\
     &\simeq&2v_K\left[\left(v_{d,\phi}-v_{g,\phi}\right)-\left(v_K-v_{g,\phi}\right)\right]\\
    &\simeq& 2 v_K[(v_{d,\phi}-v_{g,\phi})- \eta)]v_K/2
\end{eqnarray}
Solving for $v_{d,\phi}^2$ and inserting back into \autoref{eq:simple_eom_cyl}, we finally arrive at the cylindrical terminal velocity:
\begin{equation}
\label{eq:vvarpi}
v_{d,\varpi} = \frac{v_{g,\varpi}T_s^{-1} + \left[\beta\left(\frac{\varpi}{r}\right)-\eta\right] v_{\rm K}}{T_s + T_s^{-1}},
\end{equation}
where $v_{\rm K} = \varpi \Omega_{\rm K}$ is the Keplerian velocity at the position of the grain. We note that the terminal velocity approximation we have adopted breaks down as $v_{d,\varpi}$ approaches $v_{\rm K}$, and that, for $\beta > 1$, this is possible for grains whose dimensionless stopping time $T_s$ is of order unity. However, in the numerical calculations we carry out below, the largest values of $\beta$ we consider are a few, and for the grains with the largest values of $\beta$, the stopping time $T_s \ll 1$ in all cells that are included in our calculation (see \autoref{ssec:numerical_method} for details). Consequently, the maximum grain velocities we find are less than 1\% of $v_K$, in which regime \autoref{eq:vvarpi} is valid.

\subsection{Turbulent diffusion}

Having solved for the bulk velocity of the gas, we next calculate the rate of turbulent mixing. Following \citet{Takeuchi:2002}, we model the diffusive flux of dust as
\begin{equation}
\mathbf{j}_{d,a} = -\frac{1}{1+T_s^2}\alpha \frac{c_s^2}{\Omega_{\rm K}} \rho_g \nabla\left(\frac{\rho_d}{\rho_g}\right)
\equiv -D_{d,a} \rho_g \nabla f_d
\end{equation}
where $\alpha$ is the dimensionless viscosity, and we have defined $f_{d,a} = \rho_{d,a}/\rho_g$ as the dust mass fraction for grains of size $a$, and
\begin{equation}
D_{d,a} =  \alpha \frac{c_s^2}{\Omega_{\rm K}}\left(\frac{1}{1+T_s^2}\right)
\label{eq:diffcoef}
\end{equation}
as the diffusion coefficient for dust grains of size $a$. Note, however, the $D_{d,a}$ is the diffusion coefficient for grain concentration, rather than grain density.

\section{Simplified 1D System}
\label{sec:1d}

\subsection{Derivation}

Before proceeding to full numerical solution of \autoref{eq:advec_diff}, it is helpful to gain insight by considering a simplified system that we can solve semi-analytically. We do so by making the following approximations. First, we neglect the vertical structure of the disc, and focus on a small radial section so that we can neglect curvature (i.e., we treat the coordinate system as Cartesian, with the $x$ direction aligned with the radial direction), and can treat the initial dust density distribution, background gas disc, and Keplerian speed as uniform (i.e., $\rho_g$, $v_{\rm K}$, and $c_s$ are all constant).  Second, we consider only a single size of dust grain $a$, with constant stopping time $t_s$. Third, we neglect both the radial inflow of the gas and the slow inward drift of dust compared to gas as a result of drag, i.e., we set $v_{g,\varpi} = 0$ and $\eta = 0$. While these assumptions are obviously oversimplifications, they retain the essence of the problem: the radial evolution of the dust will be determined by the competition between radiation pressure forces, which attempt to sweep the dust up into an outward-moving shell, and diffusion, which attempts to force the dust distribution back towards uniform. The simplified system that we solve is very similar to that considered by \citet{Dominik11a}, though our treatment is somewhat more general in that we will not assume, as they do, that the dust is swept into an infinitely-thin wall.

Under the approximations we have described, \autoref{eq:advec_diff} reduces to the one-dimensional PDE
\begin{equation}
\label{eq:advec_diff_1d}
\frac{\partial \rho_d}{\partial t} + \left(\frac{\beta_0 v_{\rm K}}{T_s + T_s^{-1}}\right) \frac{\partial}{\partial x}\left(\rho_d e^{-\tau}\right) - D_d \frac{\partial^2 \rho_d}{\partial x^2} = 0,
\end{equation}
where we have dropped the subscript $a$'s since we are considering only a single grain size, and we orient our coordinate system so the star lies at $x\ll 0$. As an initial condition we take $\rho_d = 0$ for $x < 0$ and $\rho_d = \rho_{d,0}$ for $x > 0$, i.e., the dust initially occupies the positive half-plane. With this initial condition, we can write the optical depth to position $x$ as
\begin{equation}
\tau = \frac{3}{4 a \rho_s} \int_0^x \rho_d(y) \, dy.
\end{equation}

The first step in solving \autoref{eq:advec_diff_1d} is to non-dimensionalise it. We normalise the density to the initial density, measure length in units of the optical depth at the initial density, and measure time in units such that the diffusion coefficient is unity. Mathematically, this amounts to making a change of variables $\rho_d' = \rho_d/\rho_{d,0}$, $x'=x/x_d$, $t'=t/t_d$, where
\begin{equation}
x_d = \frac{4}{3}a \frac{\rho_s}{\rho_{d,0}}
\qquad
t_d = \frac{x_d^2}{D_d}.
\end{equation}
Here $x_d$ and $t_d$ are the characteristic length and time scales for the problem. This allows us to rewrite \autoref{eq:advec_diff_1d} as
\begin{equation}
\label{eq:advec_diff_1d_nondim}
\frac{\partial \rho_d'}{\partial t'} + \chi \frac{\partial}{\partial x'} \left(\rho_d' e^{-\tau}\right) - \frac{\partial^2 \rho_d'}{\partial x'^2} = 0,
\end{equation}
where
\begin{eqnarray}
\tau & = & \int_0^{x'} \rho_d'(y') \, dy' \\
\chi & = & \frac{4}{3} \left(\frac{\beta_0}{\alpha f_{d,0}}\right) \left(\frac{\rho_s}{\rho_g}\right)^2 \left(\frac{a}{r}\right)^2 \left(\frac{v_{\rm K}}{c_s}\right)^3.
\label{eq:chi}
\end{eqnarray}
Here we have used \autoref{eq:stopping_time} and \autoref{eq:diffcoef} for $D_d$ and $T_s$, respectively, $r = v_{\rm K}/\Omega_{\rm K}$ is the radial location of our region of interest, and $f_{d,0} = \rho_{d,0}/\rho_g$ is the initial dust fraction. The interstellar dust abundance is $f_{d,0}\approx 0.01$, but the transition discs in which we are interested have undergone considerable grain agglomeration into larger bodies, and have observed dust abundances that lie more in the range $\sim 10^{-5}-10^{-4}$ \citep[e.g.,][]{van-der-Marel16a}.

Thus we see that our simplified 1D system represents a single-parameter family of PDEs. The parameter $\chi$ characterises the relative importance of advection by radiation forces (the second term in \autoref{eq:advec_diff_1d}) and diffusion by the gas (the third term). Intuitively, we expect that radiation forces on the exposed face of the dust at $x=0$ will begin to sweep dust into an advancing wave, which will be spread out to a characteristic width determined by diffusion. The parameter $\chi$ controls the characteristic speed with which the wave moves, sweeping up dust as it goes. The value of $\chi$ in a real disc obviously varies significantly depending on the local properties, as we discuss in further detail in \autoref{ssec:1d_implications}, but for the cases of greatest interest to us we will have $\chi$ in the range tens to thousands.

\subsection{Semi-analytic Solution}

We cannot obtain an exact analytic solution even to \autoref{eq:advec_diff_1d}, but we can derive some analytic constraints on the asymptotic behaviour of the solution, which we can use to derive a semi-analytic model. First note that for material at high optical depth, i.e., any dust that begins at $x' \gg 1$, the advection term is negligible because it is proportional to $e^{-\tau}$. Thus the equation reduces to
\begin{equation}
\frac{\partial \rho_d'}{\partial t'} - \frac{\partial^2 \rho_d'}{\partial x'^2} = 0,
\end{equation}
which can we can solve via the usual similarity transformation for diffusion problems, $\zeta = x'/2\sqrt{t'}$. This reduces the problem to an ODE, which has the solution
\begin{equation}
\rho_d' = c_1 + c_2 \, \mbox{erf}\left(\zeta\right).
\end{equation}
The two constants of integration $c_1$ and $c_2$ are determined by the boundary conditions. One of them can be fixed by requiring that $\rho_d' \rightarrow 1$ as $\zeta=x'/2\sqrt{t'}\rightarrow \infty$, i.e., that the density approach the initial density far upstream of the advancing dust wave. Applying this condition, the analytic solution at large $\tau$ must approach
\begin{equation}
\rho_d' = 1 + k \, \mbox{erfc}\left(\frac{x'}{2\sqrt{t'}}\right),
\label{eq:rho_1d_upstream}
\end{equation}
where $k$ is a constant that depends on $\chi$, and $\mbox{erfc}(x) = 1 - \mbox{erf}(x)$ is the complementary error function.

Thus the solution will consist of a low-optical depth, low-density downstream region over which the dust wave has already passed, a transition zone where $\tau\approx 1$ located at position $s(t')$, and an upstream region where the solution approaches \autoref{eq:rho_1d_upstream}. At late times, when $s(t') \gg 1$, the great majority of the dust mass that was initially at $x' < s(t')$ must be in the upstream region, since by definition the downstream and transition regions contain a mass per unit area of order unity in our dimensionless variables. Thus conservation of mass requires that for $s(t')\gg 1$ we have
\begin{eqnarray}
s(t') & \approx & \int_{s(t')}^{\infty} k \, \mbox{erfc}\left(\frac{x'}{2\sqrt{t'}}\right)
\, dx' \nonumber \\
& = & k \left\{\mbox{erf}\left[\frac{s(t')}{2\sqrt{t'}}\right] - 1 + \sqrt{\frac{t'}{\pi}} \frac{e^{-s(t')^2/4t'}}{s(t')}\right\}.
\end{eqnarray}
This equation can be satisfied for arbitrary $t$ only if we have
\begin{eqnarray}
s(t') & = & 2\lambda \sqrt{t'} \\
k & = & \left[\mbox{erf}\left(\lambda\right) - 1 + \frac{e^{-\lambda^2}}{\sqrt{\pi}}\right]^{-1}.
\label{eq:k_upstream}
\end{eqnarray}
Thus we learn that the position of the dust wave at late times must be proportional to $\sqrt{t'}$, with a constant of proportionality that depends on $\chi$.\footnote{Although \autoref{eq:rho_1d_upstream} and \autoref{eq:k_upstream} are exact, they are inconvenient for practical computation when $\lambda \gtrsim 5$ because $k$ becomes very large and $\mbox{erfc}(x'/2\sqrt{t'})$ very small. In this case it is preferable to evaluate $\rho_d' = 1 + k\,\mbox{erfc}(\zeta)$ via the series expansion $\rho_d' = 1+ [2\lambda^3+3\lambda-3\lambda^{-1}+O(\lambda^{-3})]  e^{\lambda^2-\zeta^2}$, where $\zeta=x'/2\sqrt{t'}$.}

\subsection{Numerical Solution}

\begin{table}
\begin{tabular}{ccccccc}
\hline\hline
$\log\chi$ & $t_{\rm max}$ & $L$ & $\Delta x_{\rm min}$ & $k$ & $\lambda$ & $\mbox{Err}(\lambda)$ \\ \hline
1.0 & 20.0 & 16.0 & 1/32 & 1.40 & 1.96 & $7.3\times 10^{-3}$ \\
1.25 & 12.0 & 16.0 & 1/64 & 1.33 & 2.76 & $5.9\times 10^{-3}$ \\
1.5 & 8.0 & 8.0 & 1/128 & 1.32 & 3.80 & $5.0\times 10^{-3}$ \\
1.75 & 4.0 & 8.0 & 1/256 & 1.25 & 5.16 & $4.0 \times 10^{-3}$ \\
2.0 & 2.0 & 4.0 & 1/256 & 1.36 & 6.94 & $7.3 \times 10^{-3}$ \\
2.25 & 2.0 & 4.0 & 1/512 & 1.24 & 9.31 & $6.5 \times 10^{-3}$ \\
2.5 & 2.0 & 4.0 & 1/1024 & 1.13 & 12.46 & $5.5 \times 10^{-3}$ \\
2.75 & 1.6 & 4.0 & 1/1024 & 1.03 & 16.59 & $9.1\times 10^{-3}$ \\
3.0 & 1.0 & 4.0 & 1/1024 & 0.87 & 22.03 & $2.1\times 10^{-3}$ 
\\ \hline \hline
\end{tabular}
\caption{
\label{tab:1d_calculations}
Parameters and results for simulations of the simplified 1D system. Here $\chi$ is the dimensionless advection to diffusion ratio, $t_{\rm max}$ is the dimensionless time $t'$ for which we run the simulation, $L$ is the dimensionless size of the simulation domain, $\Delta x_{\rm min}$ is the spatial resolution of the best-resolved run, $k$ is the estimated order of convergence (error $\propto \Delta x^{-k}$), $\lambda$ is our best estimate for $\lambda$ based on Richardson extrapolation, and $\mbox{Err}(\lambda)$ is the estimated fractional error on $\lambda$. See main text for details of how $\Delta x_{\rm min}$, $k$, $\lambda$, and $\mbox{Err}(\lambda)$ are computed.
}
\end{table}

\begin{figure}
\includegraphics[width=\columnwidth]{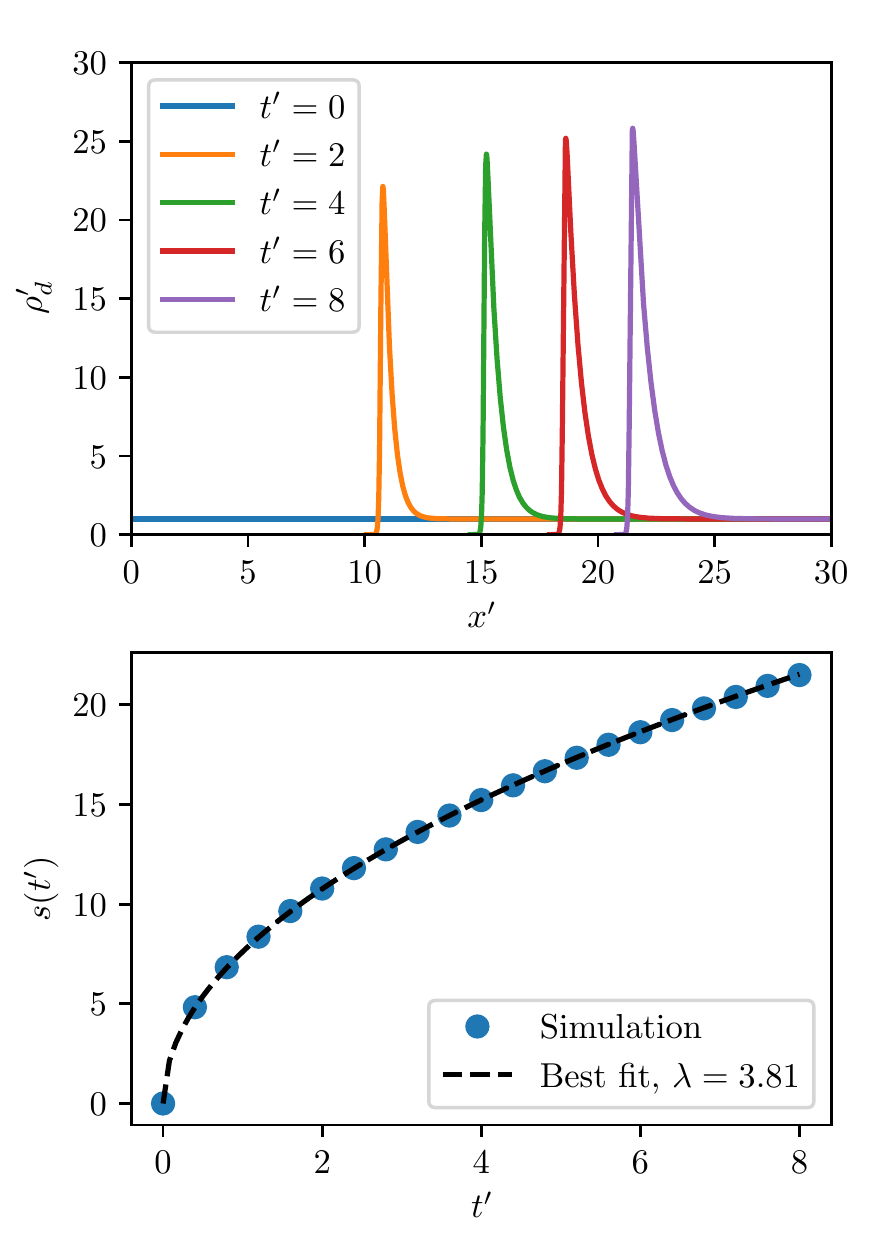}
\caption{
\label{fig:1d_example}
\textbf{Top:} dimensionless density $\rho_d'$ as a function of position $x'$ at five different times, for the case $\log\chi = 1.5$ at a resolution $\Delta x = 1/128$. Note that the region plotted is larger than the domain size $L$ due to our sliding grid; see \aref{app:numerics_1d} for details. \textbf{Bottom:} position of the dust front $s(t')$ as a function of dimensionless time $t'$. The blue points show the simulation results, where we define the front location as the position of maximum dust density (every 5th time plotted, to avoid clutter), while the black dashed line shows the best-fitting semi-analytic solution, $s(t') = 2\lambda \sqrt{t'}$.}
\end{figure}

We can verify this analytic calculation, and calibrate the dependence of $\lambda$ on $\chi$, using numerical solutions to \autoref{eq:advec_diff_1d_nondim}. We solve the system using a 1D finite volume method that is second order accurate in both space and time; we give a full description of the method in \aref{app:numerics_1d}. In \autoref{fig:1d_example} we show an example solution to \autoref{eq:advec_diff_1d_nondim} for $\log\chi = 1.5$. The parameters for this run are included in \autoref{tab:1d_calculations}. As predicted, the position of the dust front (defined here by the location of maximum dust density) as a function of time is fit extremely well by $s(t') \propto t'^{1/2}$; a simple least-squares fit to the solution shown in \autoref{fig:1d_example} gives $\lambda = 3.81$.

In order to determine the dependence of $\lambda$ on $\chi$, we solve the system numerically at a range of $\chi$ values. We list all the simulations we carry out in \autoref{tab:1d_calculations}; the domain sizes $L$ and simulation times $t_{\rm max}$ are chosen to ensure that the width of the dust wave is $\ll L$ at all times, and that the dust wave advances to $x\approx 20$, by which point the wave position as a function of time has always converged very well to the asymptotic $s\propto t'^{1/2}$ behaviour we predict analytically. We ensure that our results are converged in resolution by carrying out a convergence study: for every case, we first run the simulation at a resolution of 64 cells and then 128 cells, compute $\lambda$ via a least-squares fit to the front position as a function of time at both resolutions, and compare the results. If they differ by more than $1\%$, we double the resolution again, to 256 cells, and repeat the process. We continue doubling the resolution until either (1) we reach a resolution of 4096 cells or (2) for the highest two resolutions, the two values of $\lambda$ that emerge from our fit differ by $<1\%$. We then use a Richardson extrapolation of the resolution-dependent results to generate our final estimate for $\lambda$ for that value of $\chi$; we do this in two steps. First, we estimate the order of convergence by fitting the difference between the outcome at a given resolution at the maximum resolution as a function of resolution.\footnote{Formally our method is second-order accurate for smooth flows. However, the flow is not smooth in the vicinity of the maximum dust density, where $\tau \approx 1$. Since the behaviour in this region is critical to determining the solution, the actual accuracy will be worse than second order. We find typical convergence orders of $1-2$ depending on $\chi$.} Second, we use this estimate for the extrapolation. We list the final, extrapolated value of $\lambda$, the order of convergence, and the maximum resolution we use in \autoref{tab:1d_calculations}. We also give our estimated fractional error in $\lambda$, which we take to be the difference between the final two Richardson extrapolates, normalised by our final estimate of $\lambda$.

\begin{figure}
\includegraphics[width=\columnwidth]{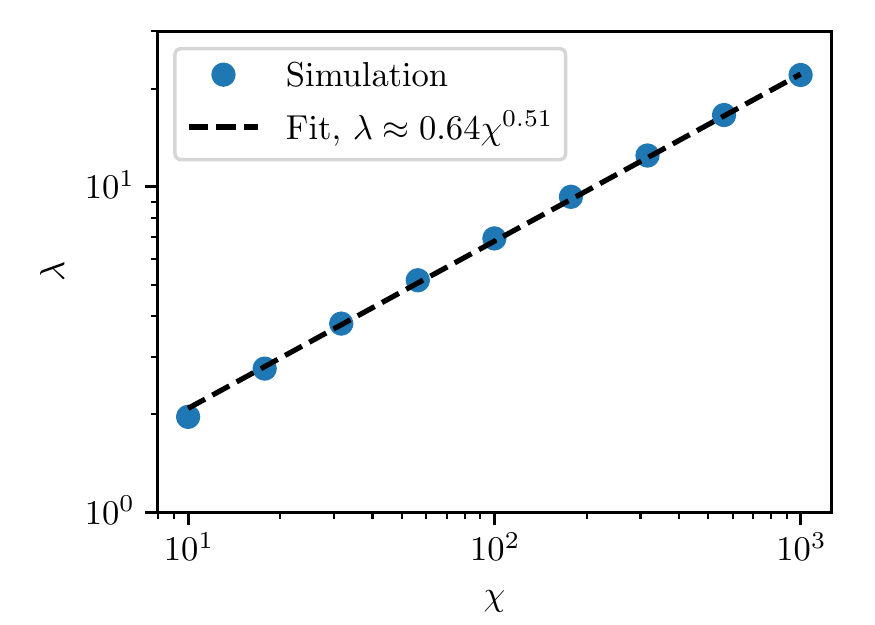}
\caption{
\label{fig:1d_lambda_vs_chi}
Dust front speed parameter $\lambda$ as a function of advection to diffusion ratio $\chi$. Blue circles show the numerical results from \autoref{tab:1d_calculations} (error bars are too small to be seen), while the black dashed line shows the best-fitting powerlaw.
}
\end{figure}

\autoref{fig:1d_lambda_vs_chi} shows $\lambda$ as a function of $\chi$ from out study. Clearly over the range we have studied, the data are consistent with a simple powerlaw relationship between $\lambda$ and $\chi$. A least squares fit to the data in \autoref{tab:1d_calculations}, including our estimated uncertainties returned by the Richardson extrapolation procedure, is
\begin{equation}
\label{eq:lambda_fit}
\lambda \approx 0.64 \chi^{0.51}.
\end{equation}

\subsection{Astrophysical Implications}
\label{ssec:1d_implications}

We are now in a position to consider the astrophysical implications of this finding. For any specified dust and gas distribution in a disc, we can compute $\chi$ from \autoref{eq:chi} at any point in the disc, and then from our semi-analytic solution $s(t') = 2\lambda \sqrt{t'}$, we can compute the characteristic time $t$ that would be required for radiation pressure to move the dust at that position a distance comparable to its current distance from the star. To be precise, at any point a radial distance $r$ from the star, we define the dimensionless radiative dust clearing time $t'$ by the condition that $r/x_d = 2 \lambda \sqrt{t'}$. The corresponding dimensional time is
\begin{equation}
t_{\rm clr} = \frac{t_d}{4\lambda^2} \left(\frac{r}{x_d}\right)^2 = \frac{1+T_s^2}{4 \alpha \lambda^2} \left(\frac{v_{\rm K}}{c_s}\right)^{2} \Omega_{\rm K}^{-1},
\label{eq:tclr1}
\end{equation}
which is the (dimensional) time that would be required for a dust wave following the semi-analytic solution derived in the previous section to move a distance $r$. For small grains, $T_s \ll 1$, in a thin, moderately-accreting disc, $v_K/c_s\sim 10^2$ and $\alpha\sim 10^{-2}$, significant dust sweeping in $\lesssim 10^6$ orbits is expected for $\lambda \sim 1-10$, corresponding to $\chi$ of tens to hundreds.

We can also write the clearing timescale in terms of the classical viscous accretion timescale $t_{\rm acc} = r^2/\nu$, where $\nu = \alpha c_s^2/\Omega_{\rm K}$ is the kinematic viscosity. Then we simply have
\begin{equation}
    t_{\rm clr} = \frac{1+T_s^2}{4\lambda^2} t_{\rm acc},
\end{equation}
and we again see that we expect grains to be cleared faster than they accrete only for $\lambda \gtrsim 1$, meaning $\chi \gtrsim 10$.

Finally, it is instructive to insert our best-fit scaling for $\lambda$ as a function of $\chi$, \autoref{eq:lambda_fit}, into \autoref{eq:tclr1}, and then to substitute for $\chi$ using \autoref{eq:chi}. Doing so gives
\begin{equation}
t_{\rm clr} \approx 0.46  \left(1+T_s^2\right) \alpha^{0.02} \left(\frac{f_{d,0}}{\beta_0}\right)^{1.02} \left(\frac{r\rho_g}{a \rho_s}\right)^{2.04} \left(\frac{c_s}{v_{\rm K}}\right)^{1.06} \Omega_{\rm K}^{-1}.
\label{eq:tclr}
\end{equation}
Thus we see that the dust clearing timescale is very sensitive to both the grain size (roughly $t_{\rm clr}\propto a^{-2}$) and the background dust density ($t_{\rm clr}\propto \rho_g^2$). Thus smaller grains in denser gas are much more resistant to clearing, while larger grains are easier to clear. This timescale is notably very insensitive to our choice of $\alpha$.
\begin{figure*}
\includegraphics[width=\textwidth]{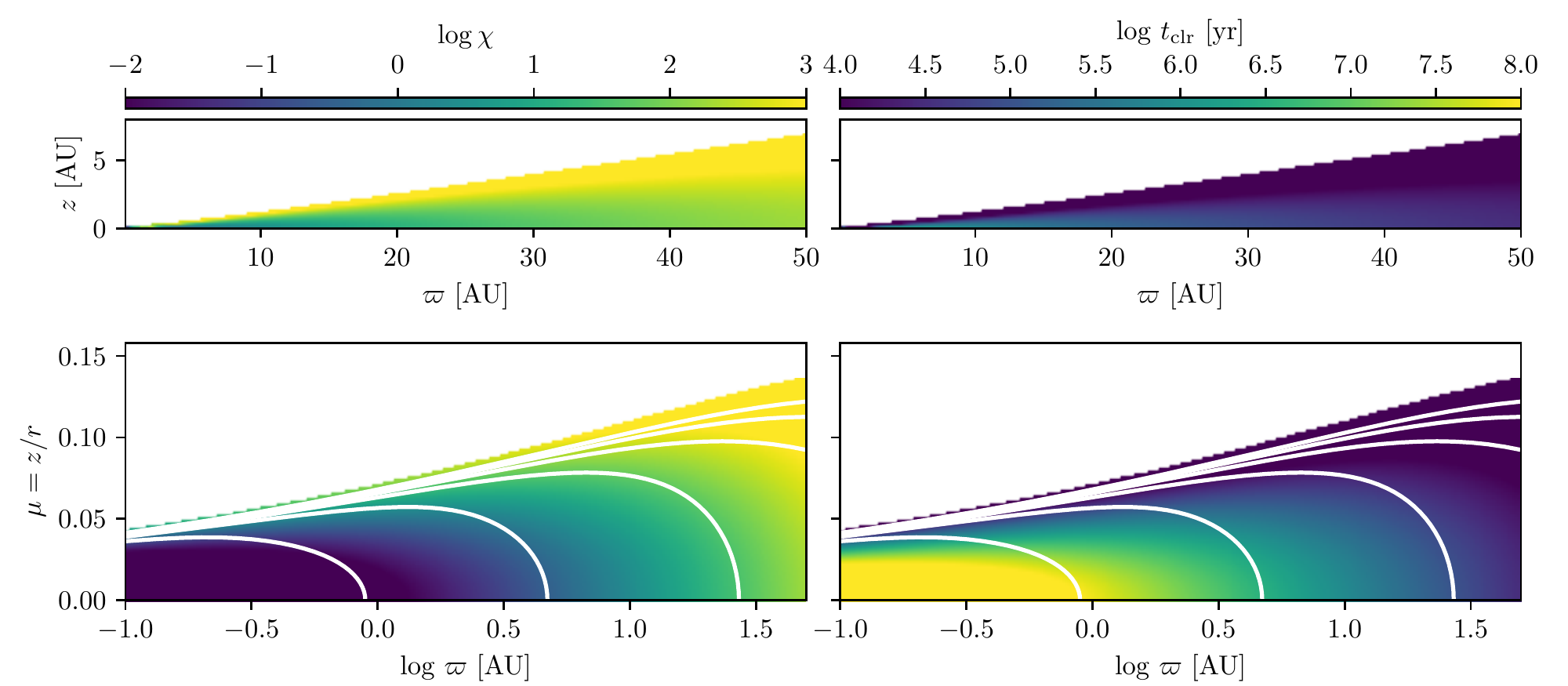}
\caption{
\label{fig:chi_tc}
Values of $\chi$ (\textbf{left}) and $t_{\rm clr}$ (\textbf{right}) in a transition disc characterised by the properties $M=M_\odot$, $L=L_\odot$, $\epsilon=0.01$, $f_{\rm dust} = 10^{-4}$, $T_0 = 120$ K, $\alpha = 10^{-3}$, $p=3/2$, $q=3/7$, $a=0.1$ $\mu$m, $\rho_s = 1$ g cm$^{-3}$. The \textbf{top} row shows these parameters in true position ($\varpi$, $z$), with the axes sized to reflect the true disc aspect ratio. In the \textbf{bottom} row we show the same data, but in coordinates ($\log\varpi$, $\mu = z/r$), so that the inner disc is more clearly visible, and radial rays from the central star correspond to horizontal lines. White lines are contours of dust density $\rho_d$, starting with $\rho_d = 10^{-24}$ g cm$^{-3}$ for the lowest contour and increasing by factors of 100 for each successive contour. We do not show $\chi$ and $t_{\rm clr}$ for $\rho_d < 10^{-30}$ g cm$^{-3}$.
}
\end{figure*}

To give a sense of the numerical values implied by \autoref{eq:tclr}, let us consider a disc in which the dust and gas are initially in equilibrium in the absence of radiative forces or radial transport (i.e., with $v_{d,a,\varpi} = 0$, $j_{d,a,\varpi} = 0$, and $\beta = 0$ in \autoref{eq:advec_diff}). \citet{Takeuchi:2002} show that the steady-state vertical distribution of size $a$ in such a disc has a steady-state solution
\begin{equation}
\rho_{d,a}(z) = \rho_{d,a}(0) \exp\left\{-\frac{z^2}{2h_g^2} - \frac{\mathrm{Sc}\, T_{s,\rm mid}}{\alpha} \left[\exp\left(\frac{z^2}{2h_g^2}\right)-1\right]\right\},
\label{eq:TL_solution}
\end{equation}
where $T_{s,\rm mid}$ is the stopping time evaluated at the midplane gas density, $\mathrm{Sc} = 1 + T_{s,\rm mid}^2$ is the Schmidt number for a particular grain size $a$, and the midplane density $\rho_{d,a}(0)$ is set by requiring that 
\begin{equation}
2 \int_0^\infty \rho_{d,a}\, dz = f_{\rm dust} \Sigma_g.
\end{equation}
Here $\Sigma_g$ is the gas column density at cylindrical radius $\varpi$, and $f_{\rm dust}$ is the dust to gas mass ratio for grains of the considered size. For any given choices of parameters describing the star ($M$, $L$) and the disc ($\epsilon$, $f_{\rm dust}$, $T_0$, $\alpha$, $p$, $q$) and the dust ($a$, $\rho_s$), we can use this expression to compute the dust and gas densities at every point, and then use these to compute $\chi$ and $t_{\rm clr}$. 

In \autoref{fig:chi_tc} we show an example map of $\chi$ and $t_{\rm clr}$ for a transition disc with $\epsilon=10^{-2}$, $f_{\rm dust} = 10^{-4}$. In the example shown, the midplane is quite resistant to dust clearing ($t_{\rm clr} \gg 1$ Myr, but above the midplane significant clearing is possible on timescales well under a Myr. Recall, however, that we have the near-proportionality $t_{\rm clr}\propto f_{d,0} \rho_g^2$. Thus we expect clearing to become much more rapid as we move to discs that are more dust- or gas-depleted than the example shown in \autoref{fig:chi_tc}. Conversely, for richer discs clearing will be much slower. In this context, it is interesting to compare our results to those of \citet{Dominik11a}, who also investigated the fate of an advancing dust wall driven outward by radiation. They concluded that radiation pressure would be ineffective, because the dust wall would quickly slow to tiny velocities that would easily be overwhelmed by even a small amount of inward gas accretion. However, this is not surprising, because they considered discs with properties close to the MMSN ($\epsilon\sim 1$, $f_{\rm dust} \sim 10^{-2}$); in terms of the dimensionless parameters, such discs have values of $\chi$ that are smaller than those shown in \autoref{fig:chi_tc} by a factor of $\sim 10^4$, and values of $t_{\rm clr}$ that are larger by $\sim 10^2$. Thus their conclusion that radiation pressure is ineffective in such a disc is entirely consistent with our finding that radiative clearing is effective only in discs that are significantly depleted in both gas and dust compared to the MMSN.

\section{2D Simulations}
\label{sec:2D}
Armed with the general understanding provided by the simplified 1D system solved in \autoref{sec:1d}, we now proceed with full numerical solutions to \autoref{eq:advec_diff} in 2D, including the full spatial dependence of the background gas disc. We summarise the properties of the runs we carry out in \autoref{tab:2d_params}. Motivated by \autoref{fig:chi_tc}, we take $\epsilon=10^{-2}$, $f_{\rm dust}=10^{-4}$ as our most gas- and dust-rich case, and explore from those values downward. Note that for an evolved disk, $f_{\rm dust}$ does {\em not} represent the total dust to gas ratio, only the dust to gas ratio for those grains that fall into the size range we simulate ($0.18 - 18$ $\mu$m); we omit larger grains, since these are expected to have settled to the midplane by the transition phase, and radiative effects are unimportant for them in any event. Similarly, \citet{Ercolano17a} find that observed transition discs span accretion rates from $\sim 10^{-8}$ $M_\odot$ yr$^{-1}$ down to $< 10^{-11}$ $M_\odot$ yr$^{-1}$. We are interested in discs nearing the ends of their lives, so we explore laminar accretion rates at the moderate to low end of this range, $10^{-9}$ $M_\odot$ yr$^{-1}$ and down.
Note that while we refer to runs without laminar flow as ``non-accreting", our chosen $\alpha$ does produce minimal accretion,  $\dot{M} \lesssim 10^{-12}M_\odot/{\rm{yr}}^{-1}$.

\begin{table*}
    \centering
    \begin{tabular}{cccccccccccccc}
    \hline\hline
        Parameter & Meaning & \multicolumn{12}{c}{Value} \\ \hline
        Fixed values \\ \hline
        $r_{\rm min},r_{\rm max}$ & Inner, outer radii & \multicolumn{12}{c}{0.1, 50 AU} \\
        $\mu_{\rm max} = \sin\theta_{\rm max}$ & Angular grid extent & \multicolumn{12}{c}{0.1} \\
        $N_r\times N_\theta$ & Grid resolution & \multicolumn{12}{c}{$512\times 256$} \\
        $N_a$ & \# grain size bins & \multicolumn{12}{c}{4} \\
        $\log(a_{1,2,3,4}/\mu\mathrm{m})$ & Grain sizes & \multicolumn{12}{c}{$-0.5$, 0, $0.5$, $1$} \\ 
        $M$ & Stellar mass & \multicolumn{12}{c}{1 $M_\odot$} \\
        $L$ & Stellar luminosity & \multicolumn{12}{c}{1 $L_\odot$} \\
        $q$ & Temperature index & \multicolumn{12}{c}{$-3/7$} \\
        $T_0$ & Temperature at 1 AU & \multicolumn{12}{c}{120 K} \\
        $\alpha$ & Viscosity parameter & \multicolumn{12}{c}{$10^{-4}$} \\
        $\rho_s$ & Grain density & \multicolumn{12}{c}{1.0 g cm$^{-3}$} \\
        $q_d$ & Grain size index & \multicolumn{12}{c}{$-11/6$} \\
        \hline
        Variable values & \multicolumn{1}{r}{Case:} & A1 & A2 & A3 & B1 & B2 & B3 & C1 & C2 & C3 & D1 & D2 & D3 \\ \hline
        $p$ & Gas density index & $-1.5$ & $-1.5$ & $-1.5$ & $0$ & $0$ & $0$ & $-1.5$ & $-1.5$ & $-1.5$ & $0$ & $0$ & $0$ \\
        $\log\epsilon$ & $\Sigma/\Sigma_{\rm MMSN}$ at 1 AU & $-2$ & $-2$ & $-3$ & $-2.5$ & $-2.5$ & $-2.5$ & $-2$ & $-2$ & $-2$ & $-2.5$ & $-2.5$ & $-2.5$\\ 
        $\log f_{\rm dust}$ & Initial D/G ratio & $-4$ & $-5$ & $-4$ & $-3$ & $-4$ & $-5$ & $-5$ & $-5$ & $-5$ & $-4$ & $-4$ & $-4$\\
        $\log (\dot{M}/M_\odot\,\mathrm{yr}^{-1})$ & Laminar inflow rate & - & - & - & - & - & - & $-9$
        & $-10$ & $-11$ & $-9$
        & $-10$ & $-11$
        \\
        \hline\hline
    \end{tabular}
    \caption{Summary of 2D simulation parameters}
    \label{tab:2d_params}
\end{table*}

\subsection{Numerical Method}
\label{ssec:numerical_method}

We solve \autoref{eq:advec_diff} using a conservative finite volume method that we fully describe in \aref{app:numerics}. Our method is second-order accurate in time, second-order accurate in space for the diffusion terms, and third-order accurate for the advection terms. The calculation operates on a 2D spherical polar grid defined by coordinates $(r,\theta)$, which we divide in $N_r \times N_\theta$ cells. For convenience, since we will go back and forth between polar and cylindrical coordinates, we will use $\mu = \sin\theta=z/r$ as our coordinate rather than $\theta$; $z$ and $\mu$ both increase in the same direction, and $\mu=0$ corresponds to $z=0$. The inner and outer radial edges of the grid lie at $r = r_{\rm min}$ and $r_{\rm max}$, respectively, and in the polar direction the edges of the outermost cells are at $\mu = 0$ and $\mu = \mu_{\rm max}$. We assume symmetry about the midplane at $\mu=0$. All the simulations we present here use $N_r=512$, $N_\theta=256$, $r_{\rm min} = 0.1$ AU, $r_{\rm max} = 50$ AU, and $\mu_{\rm max} = 0.1$.

In each computational cell we track the density of $N_a$ logarithmically-spaced grain size bins, each with mean grain size $a_k$. That is, the density of grains in size bin $k$ represents the total density of grains with sizes from $\sqrt{a_{k-1}a_k}$ to $\sqrt{a_k a_{k+1}}$, where $k = 1\ldots N_a$, and $a_0$ and $a_{N_a+1}$ are set so that bins 1 and $N_a$ contain the same logarithmic range of grain sizes as all other bins. For all simulations we present here, we adopt $N_a = 4$ with $a_k = 10^{(k-1)/2}$ $\mu$m, so mean grain sizes go from $10^{-0.5} - 10$ $\mu$m in steps of 0.5 dex, and grain size bins go from $10^{-0.75} - 10^{1.25}$ $\mu$m, with bins 0.5 dex wide.

We adopt zero flux boundary conditions across both the midplane at $\theta = 0$ and the top of the disc at $\theta = \theta_{\rm min}$. At the inner and outer radial boundaries of our computational domain we adopt diode boundary conditions. At the inner edge we set the diffusive mass flux to zero, and we set the mass flux across the boundary to zero in any location where the velocity is into the computational domain, but we allow mass to flow inward radially if the velocity at the domain edge is inward. For the outer boundary, we similarly set the diffusive mass flux to zero, and anywhere the radial velocity is out of the domain we allow mass to flow out freely, but no mass to enter.

\subsection{Initial Conditions}

We initialise all simulations using the analytic solution derived by \citet{Takeuchi:2002}, given by \autoref{eq:TL_solution}. However, unlike in \autoref{sec:1d}, we now consider multiple grain size bins, and thus the constraint on the initial midplane density $\rho_{d,a}(0)$ becomes
\begin{equation}
    2 \int_0^\infty \rho_{d,a} \, dz = f_{\rm dust} f_a \Sigma_g,
\end{equation}
where $\Sigma_g$ is the gas column density at cylindrical radius $\varpi$, $f_{\rm dust}$ is the initial total dust to gas ratio summed over all grain sizes, and $f_a$ is the fraction of the total grain mass found in grains in the size bin whose mean size is $a$. We set the initial fractional masses $f_a$ in each grain size bin by assuming that grains follow a size distribution consistent with a collisional cascade, $dn/dm \propto m^{q_d}$, where $m \propto a^3$ is the mass of an individual grain and $q_d \approx -11/6$, as expected for a collisional cascade after larger bodies have started to form \citep{Dohnanyi69a}. From this choice, we have $f_{a,k} \propto (a_{k+1}a_k)^{3(2+q_d)/2} - (a_{k}a_{k-1})^{3(2+q_d)/2}$ for size bin $k$, which together with the constraint $\sum_k f_{a,k} = 1$ fully specifies the initial mass in each bin.

We carry out twelve simulations, broken into four series of three each, using the parameters shown in \autoref{tab:2d_params}. The first series, A1 - A3, uses our fiducial $p=-1.5$ density profile for the initial gas and dust disc \citep{ChiangYoudrev09}, with varying amounts of gas and dust depletion. Observational 
disc modelling gives a range of density profiles from flat through to $p=-1.5$, with a mean around $p=-0.9$ \citep{Williams:2011}, so our series A runs are at the centrally-concentrated end of the observed range.
The most gas- and dust-rich of these cases (case 1 in the table, with $\epsilon=10^{-2}$, $f_{\rm dust}=10^{-4}$) corresponds to the parameters shown in \autoref{fig:chi_tc}, and roughly to the highest column density inner transition disc in the sample of \citet{van-der-Marel16a}; case 2 has a lower initial dust-to-gas ratio by a factor of 10, while case 3 has the same dust-to-gas ratio as case 1, but a factor of 10 lower disc mass overall. The series A simulations have no laminar inflow, and thus are appropriate to represent transition discs that have nearly ceased gaseous accretion. They are intended to explore how the results depend on the properties of the disc in the case where the disc is centrally-concentrated but non-accreting. Note that the total disc mass modelled in the series A runs is significantly lower than inferred for typical transition disks \citep{Najita15} based on continuum mm observations, a fixed opacity, and dust to gas mass ratios based on an unevolved disc. Although the purpose of these models is to describe a plausible mid-point between known disks and cleared disks (i.e., while ``transition''-ing), this fact nonetheless inspires a series of models with less steep density profile power laws.

The next series of three simulations, B1 - B3, uses a flat gas density profile $p=0$, as might be expected to prevail late in disc evolution after the majority of the gas in an inner disc accretes onto the host star. These are interesting both because they potentially represent the evolution at such late stages of accretion, and because the choice $p=0$ implies that there is no radial drift of grains into the star, and thus no other grain removal mechanism operates. For these cases we fix the ratio of mass relative to the MMSN at 1 AU to $\epsilon=10^{-2.5}$, and use three dust abundances $f_{\rm dust} = 10^{-3}$, $10^{-4}$, and $10^{-5}$; note that, since the gas surface density is normalised at 1 AU, our choice $\epsilon=10^{-2.5}$ implies that, at the outer disc edge at 50 AU, the gas surface density is in fact 10\% larger than that of the fiducial MMSN at that radius. Furthermore, the total gas mass in these discs is $\sim$6\,M$_\text{J}$, which is typical of observe transition disks \citep{Najita15}. As with the first three models, these cases have zero laminar accretion rate.

The remaining two simulation series, C and D, use initial conditions identical to those in cases A2 ($p = -1.5$, $\epsilon = 10^{-2}$, $f_{\rm dust} = 10^{-5}$) and B2 ($p=0$, $\epsilon = 10^{-2.5}$, $f_{\rm dust} = 10^{-4}$), respectively, but also include laminar inflow rates of $10^{-9}$, $10^{-10}$, and $10^{-11}$ $M_\odot$ yr$^{-1}$. Thus in these series we consider our intermediate cases of centrally-concentrated and flat-profile discs, and probe how the results change over the moderate to low range of accretion rates found by \citet{Ercolano17a}.

\subsection{Simulation results}

\subsubsection{Non-accreting cases with a radial gas gradient}

\begin{figure}
    \centering
    \includegraphics[width=\columnwidth]{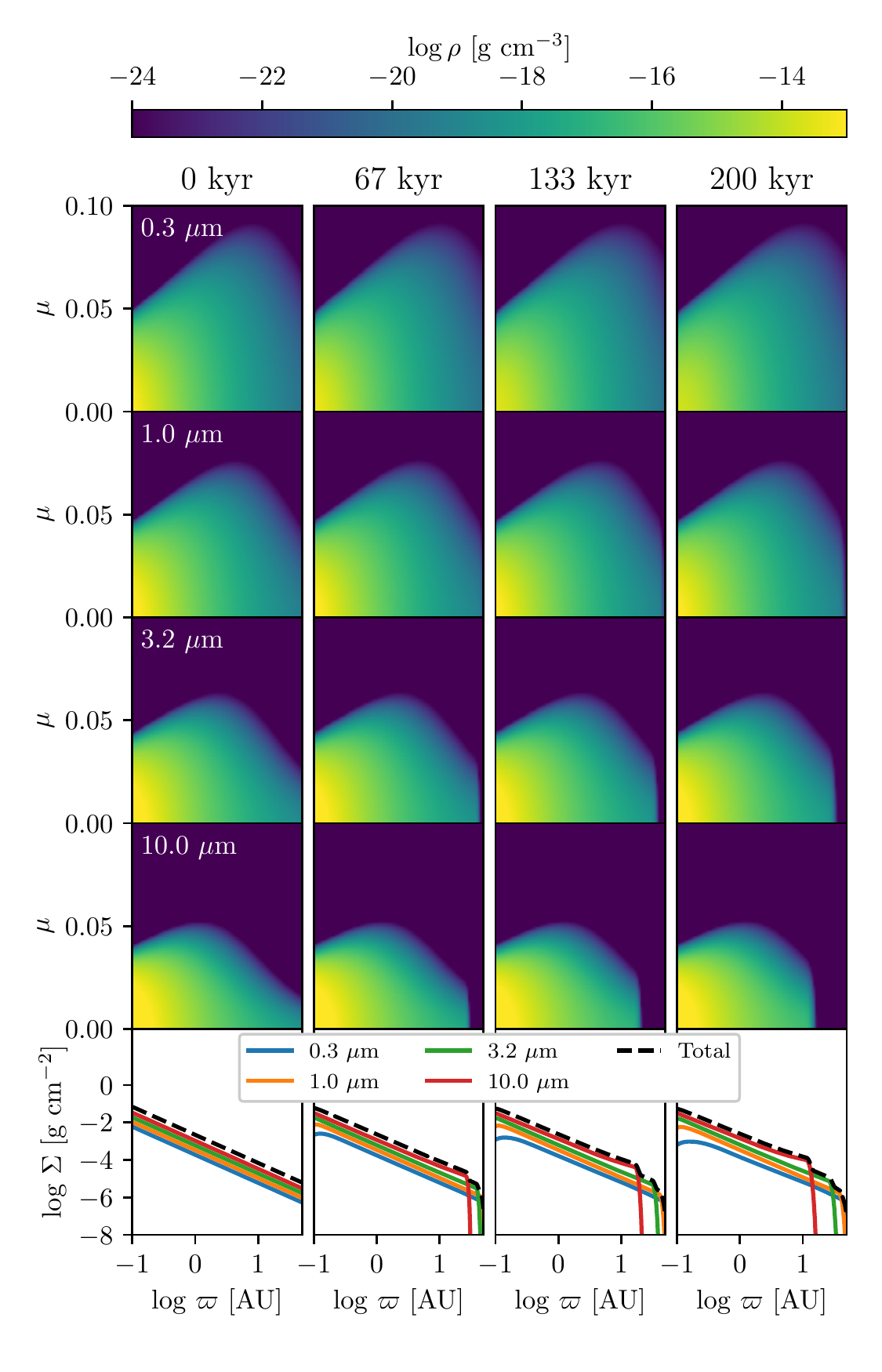}
    \caption{Snapshots of the dust distribution in our simulation with $\epsilon=10^{-2}$, $f_{\rm dust}=10^{-4}$, and $p=-1.5$ (case A1 in \autoref{tab:2d_params}). Each column shows the distribution at a different time, as indicated at the top of the column. The top four rows show the density $\rho_{d,a}$ of grains of radius $a=10^{-0.5}, 1, 10^{0.5}, 10$ $\mu$m, as indicated by the white labels on each row, as a function of position; we use $\log \varpi$ and $\mu=\sin\theta$ as our position coordinates, so in this projection radial rays from the star correspond to horizontal lines. The bottom row shows the vertically-integrated column density $\Sigma_d$ for each grain size bin, and summed over all grain sizes (see legend), as a function of log radius. An animated version of this figure is included in the Supplementary material (online).}
    \label{fig:eps2_fd4_slices}
\end{figure}

\begin{figure}
    \centering
    \includegraphics[width=\columnwidth]{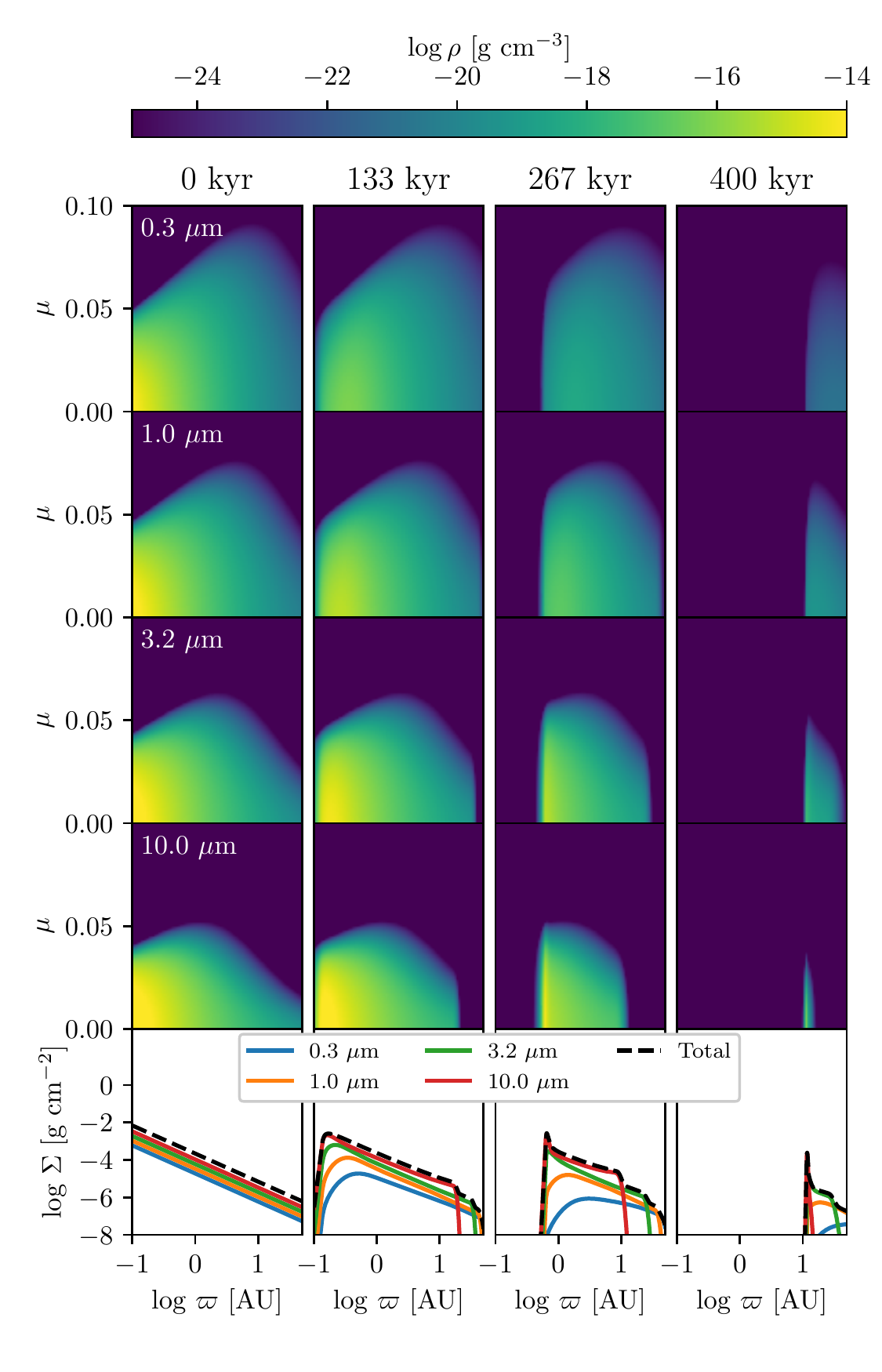}
    \caption{Same as \autoref{fig:eps2_fd4_slices}, but for the run with $\epsilon=10^{-2}$, $f_{\rm dust} = 10^{-5}$, $p=-1.5$ (case A2 in \autoref{tab:2d_params}); note that the colour scales in the two figures are not the same. Due to the higher $\chi$ value, the dust is efficiently pushed outwards on timescales of 100 kyr. An animated version of this figure is included in the Supplementary material (online).}
    \label{fig:eps2_fd5_slices}
\end{figure}

\begin{figure}
    \centering
    \includegraphics[width=\columnwidth]{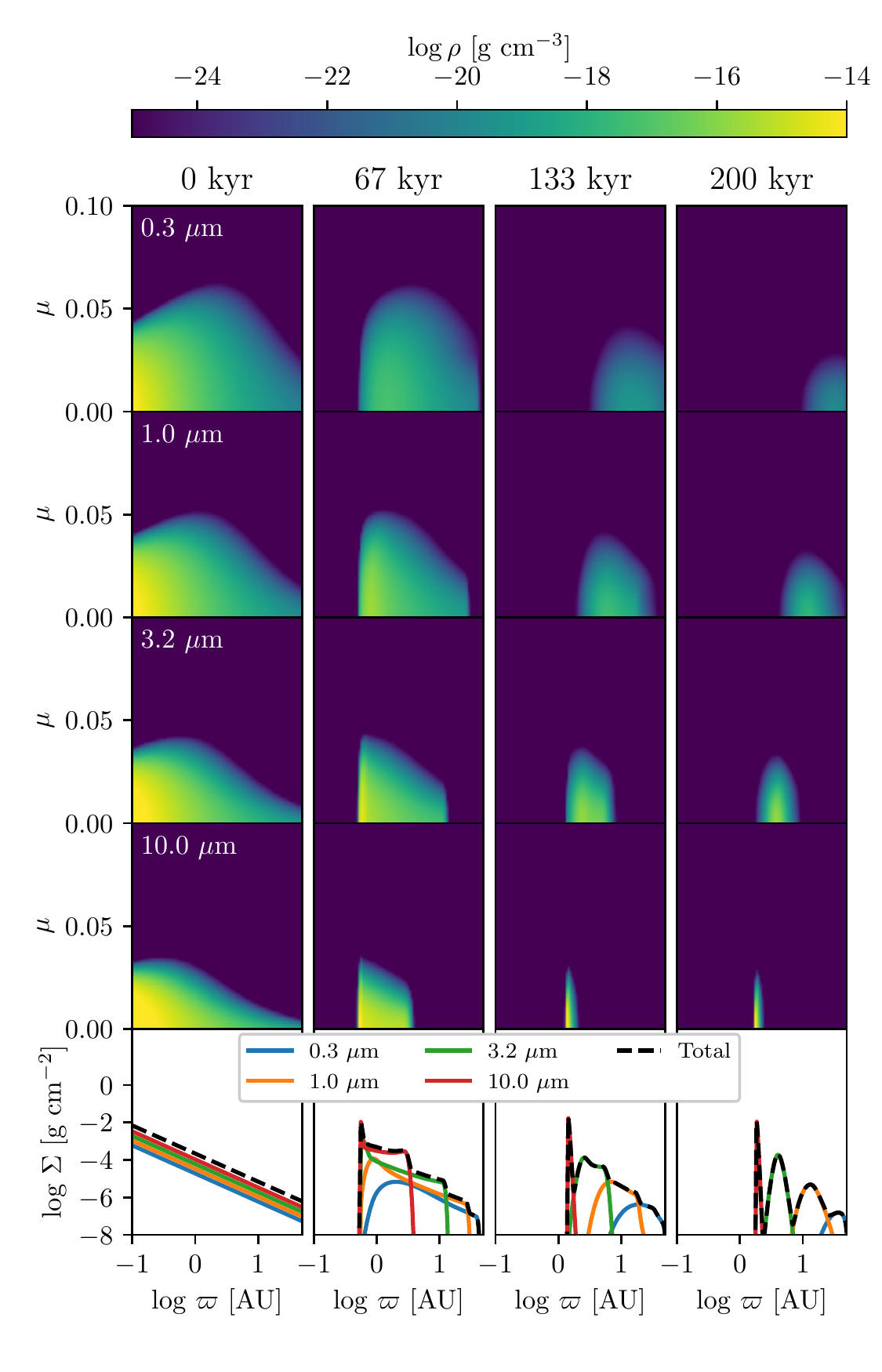}
    \caption{Same as \autoref{fig:eps2_fd4_slices}, but for the run with $\epsilon=10^{-3}$, $f_{\rm dust} = 10^{-4}$, $p=-1.5$ (case A3 in \autoref{tab:2d_params}); note that the colour scales in the two figures are not the same. The order of magnitude reduction in gas density compared to case 2 facilitates the collection of dust into rings, whose location and shape are grain-size dependent. An animated version of this figure is included in the Supplementary material (online).}
    \label{fig:eps3_fd4_slices}
\end{figure}

We show snapshots of our simulation results for cases A1 - A3 in \autoref{fig:eps2_fd4_slices}, \autoref{fig:eps2_fd5_slices}, and \autoref{fig:eps3_fd4_slices}, respectively. Comparing the runs, we see some common features and some differences. As expected, our initial distribution of grains places the largest grains at the smallest scale height. Because of their small scale height, the largest grains at the outer edge of the disc are completely shielded from radiation by the dense inner disc. As a result the grains drift inward from the outer edge of the disc at 50 AU, leaving a void behind them; this is the usual result of gas drag, and is fastest for the largest grains because they are nearest to be being critically-damped, $T_s \approx 1$. 
At smaller radii, where grains are exposed to radiation pressure, the situation is very different. In case A1, which is our most gas- and dust-rich, the inner disc is close to static over the duration of our simulation. This is simply a consequence of the small values of $\chi$ for the inner part of the disc shown in \autoref{fig:chi_tc}: due to the strong drag forces imposed by high gas densities, the net rate of grain drift is small. The smallest grains that are lofted well above the disc plane and that are exposed to radiation can drift at appreciable speeds, but the mass of grains in regions that are subject to drift is negligible compared to the much larger mass in regions where drift is negligible. Consequently any grains that are pushed outward by radiation are immediately replaced as turbulence causes the much larger reservoir of low-altitude grains to diffuse upward.

In cases A2 and A3, on the other hand, the outcome is very different. Case A2 has lower shielding against radiation due to its lower dust mass, while case A3 has both lower shielding and reduced drag. Both lead to a substantially higher value of $\chi$, and a shorter dust clearing time, such that dust is driven back from the disc inner edge on timescales of $\sim 100$ kyr. The smallest grains are swept up most rapidly, because their greater height within the disc leaves them both more exposed to stellar radiation, and less slowed by gas drag. Larger grains that sit lower in the disc move outward more slowly, and form a sharper ring of dust due to the stronger drag forces in the regions where they reside. Consequently, radiation sorts the grains by size; this sorting is especially apparent for case A3 (\autoref{fig:eps3_fd4_slices}). However, the entire structure of sorted grains moves outward over time, eventually colliding with the outer edge of shielded grains drifting inward due to gas drag. At this point all the grains are collected into rings that radiation pushes outward. If we allow the simulation to run long enough, eventually all the dust leaves the computational domain.

\subsubsection{Non-accreting cases without a radial gas gradient}

\begin{figure}
    \centering
    \includegraphics[width=\columnwidth]{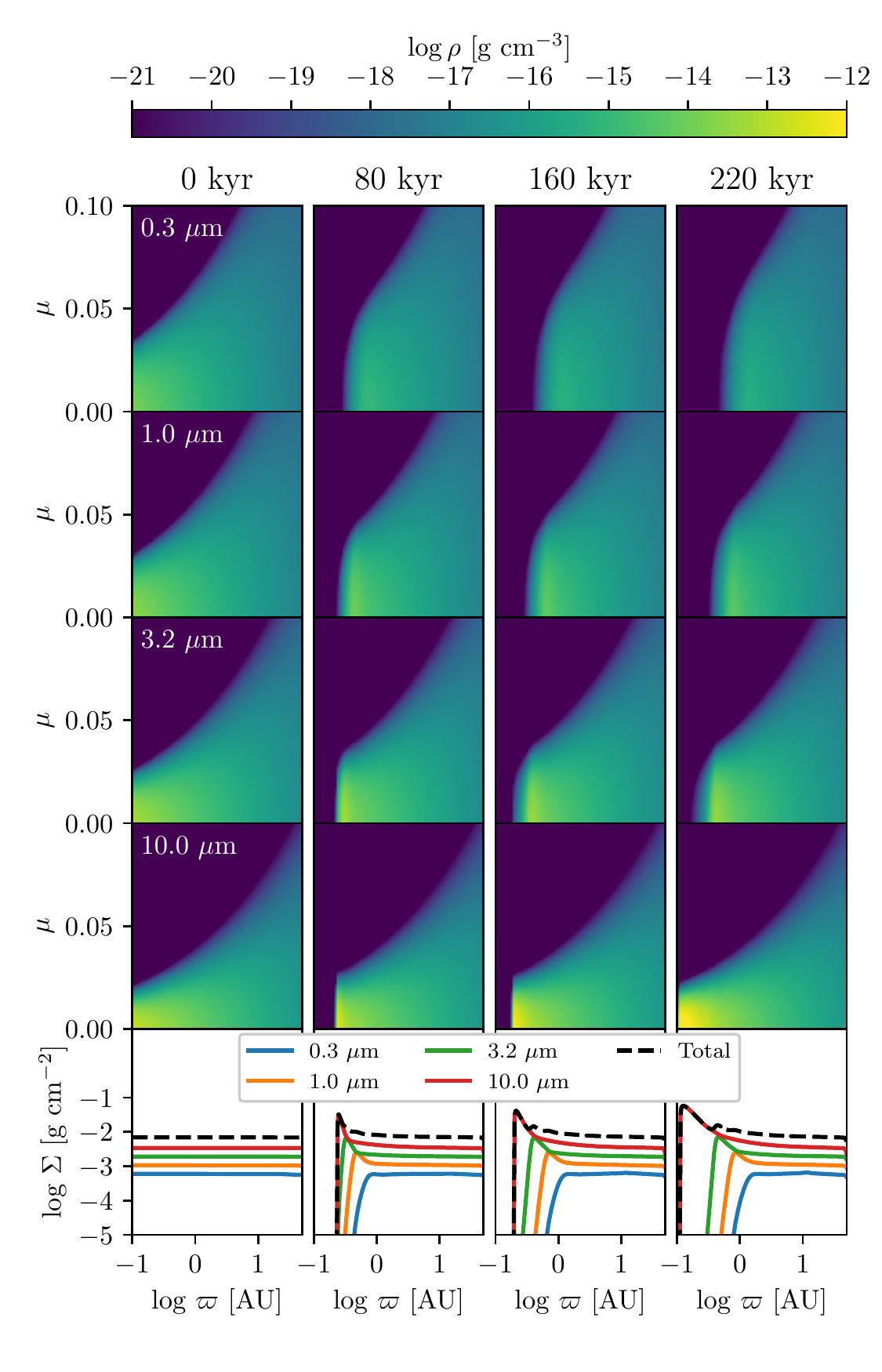}
    \caption{Same as \autoref{fig:eps2_fd4_slices}, but for the run with $\epsilon=10^{-2.5}$, $f_{\rm dust} = 10^{-3}$, $p=0$ (case B1 in \autoref{tab:2d_params}); note that the colour scales in the two figures are not the same. The absence of radial density (and pressure) gradients inhibits the formation of rings and an inner hole when the gas mass is non-negligible. An animated version of this figure is included in the Supplementary material (online).}
    \label{fig:eps25p0_fd3_slices}
\end{figure}

\begin{figure}
    \centering
    \includegraphics[width=\columnwidth]{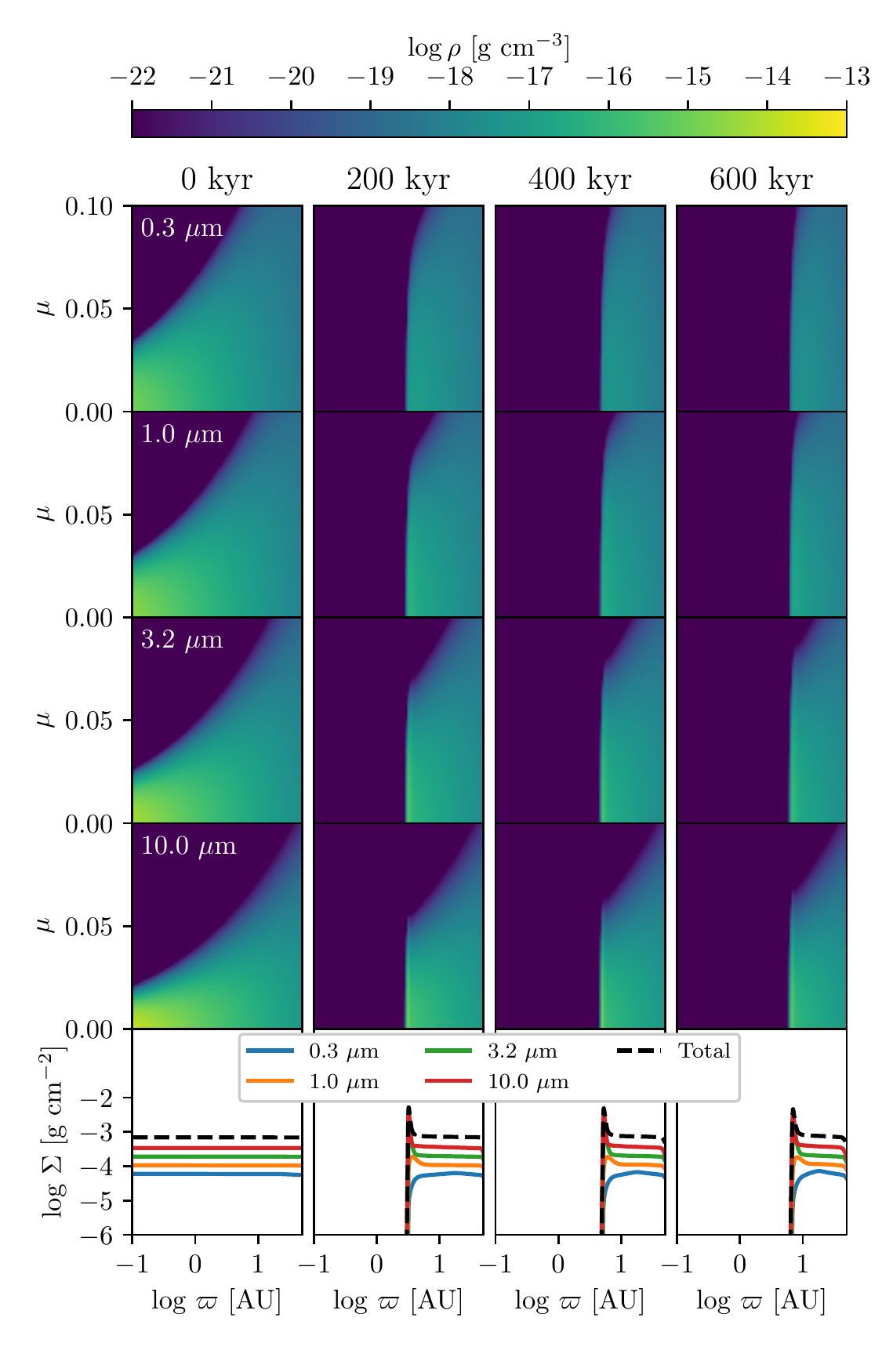}
    \caption{Same as \autoref{fig:eps2_fd4_slices}, but for the run with $\epsilon=10^{-2.5}$, $f_{\rm dust} = 10^{-4}$, $p=0$ (case B2 in \autoref{tab:2d_params}); note that the colour scales in the two figures are not the same. The timescale for clearing the outer disk increases markedly when compared to similar systems (cases A1 and A3) with a radial pressure gradient. An animated version of this figure is included in the Supplementary material (online).}
    \label{fig:eps25p0_fd4_slices}
\end{figure}

\begin{figure}
    \centering
    \includegraphics[width=\columnwidth]{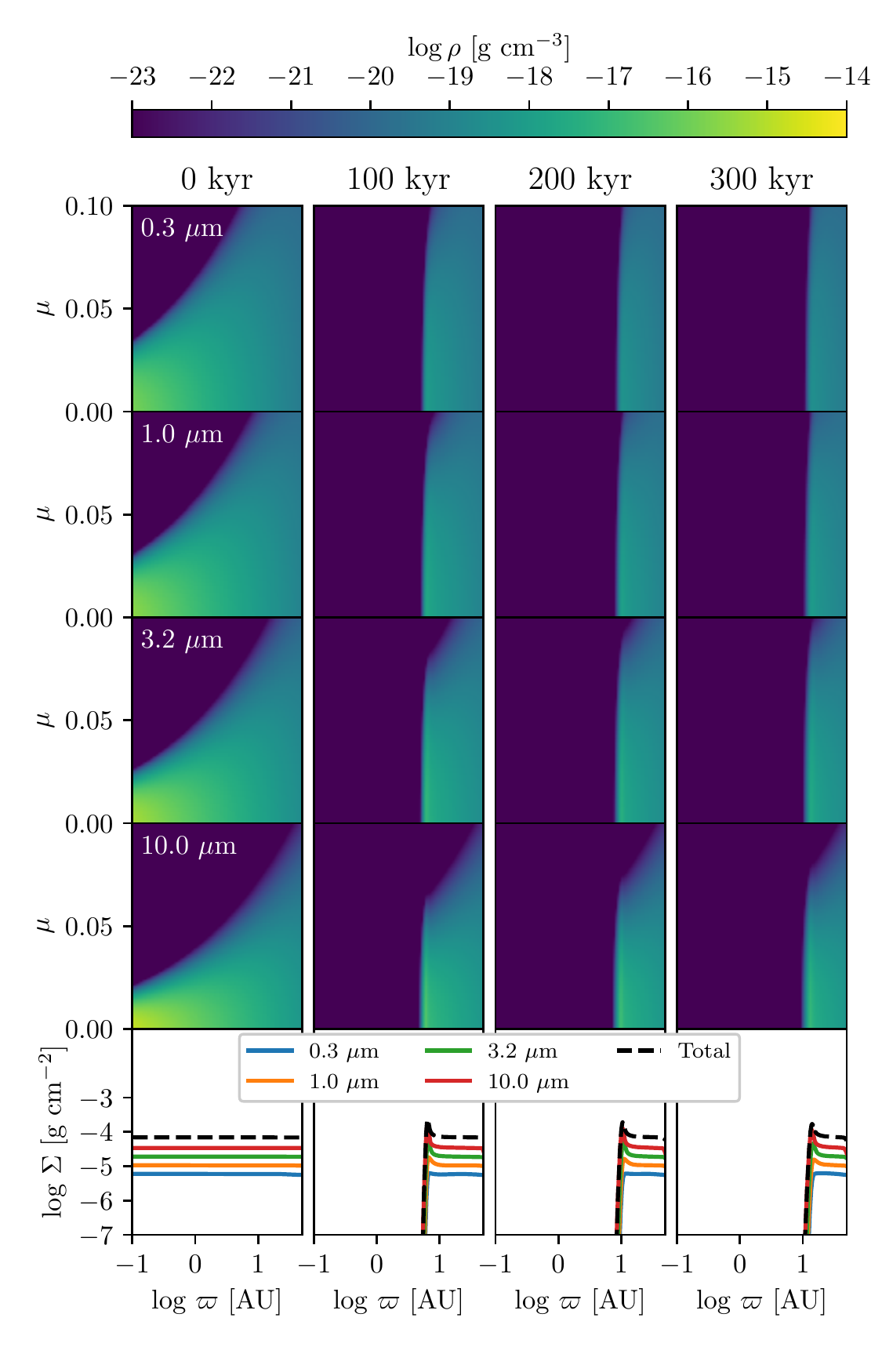}
    \caption{Same as \autoref{fig:eps2_fd4_slices}, but for the run with $\epsilon=10^{-2.5}$, $f_{\rm dust} = 10^{-5}$, $p=0$ (case B3 in \autoref{tab:2d_params}); note that the colour scales in the two figures are not the same. The absence of a radial pressure gradient inhibits ring formation because grains do not experience inward drift. Instead a wide and long-lived inner hole forms. An animated version of this figure is included in the Supplementary material (online).}
    \label{fig:eps25p0_fd5_slices}
\end{figure}

In \autoref{fig:eps25p0_fd3_slices}, \autoref{fig:eps25p0_fd4_slices}, and \autoref{fig:eps25p0_fd5_slices} we show the results for cases B1 - B3, respectively, our three cases without an initial radial gradient in the gas or dust surface density. These runs show a qualitatively different evolution from the previous cases, in that there is no inward migration of dust caused by drag. Instead, there is only outward flow of the dust caused by radiation pressure, which sweeps up an outward-moving front. Grains sort by size, but by a smaller amount than in the cases with $p=-1.5$.

Moreover, the radius of the front versus time is quite different than in series A. In the cases with $p=-1.5$, the most difficult part of the disc to evacuate is the centre. Once the central regions are clear, however, the process tends to run away: the declining density with radius, and thus the decrease in both mass to be swept up and strength of diffusive mixing, makes it relatively easy to clear the entire disc inside a few hundred kyr. For the cases with $p=0$, neither the amount of material nor the strength of diffusive mixing decrease with radius, and thus the inner part of the disc becomes easier to clear than the outer part. In the cases with $f_{\rm dust} = 10^{-4}$ and $10^{-5}$ (\autoref{fig:eps25p0_fd4_slices}, and \autoref{fig:eps25p0_fd5_slices}), the inner 1 AU of the disc is evacuated in only $\sim 50$ kyr, but the front does not reach 10 AU until hundreds of kyr. Thus in this configuration we expect to find a long-lived dust hole in discs. By contrast in the case with $f_{\rm dust} = 10^{-3}$, the smallest grains are only evacuated to $\sim 1$ AU after 200 kyr, while the largest grains stall: after moving outward for $\approx 150$ kyr, they cease to be pushed back from the star, and after some time even begin to re-occupy the central region from which they were at first expelled.

\subsubsection{Cases with laminar inflow}

\begin{figure}
    \centering
    \includegraphics[width=\columnwidth]{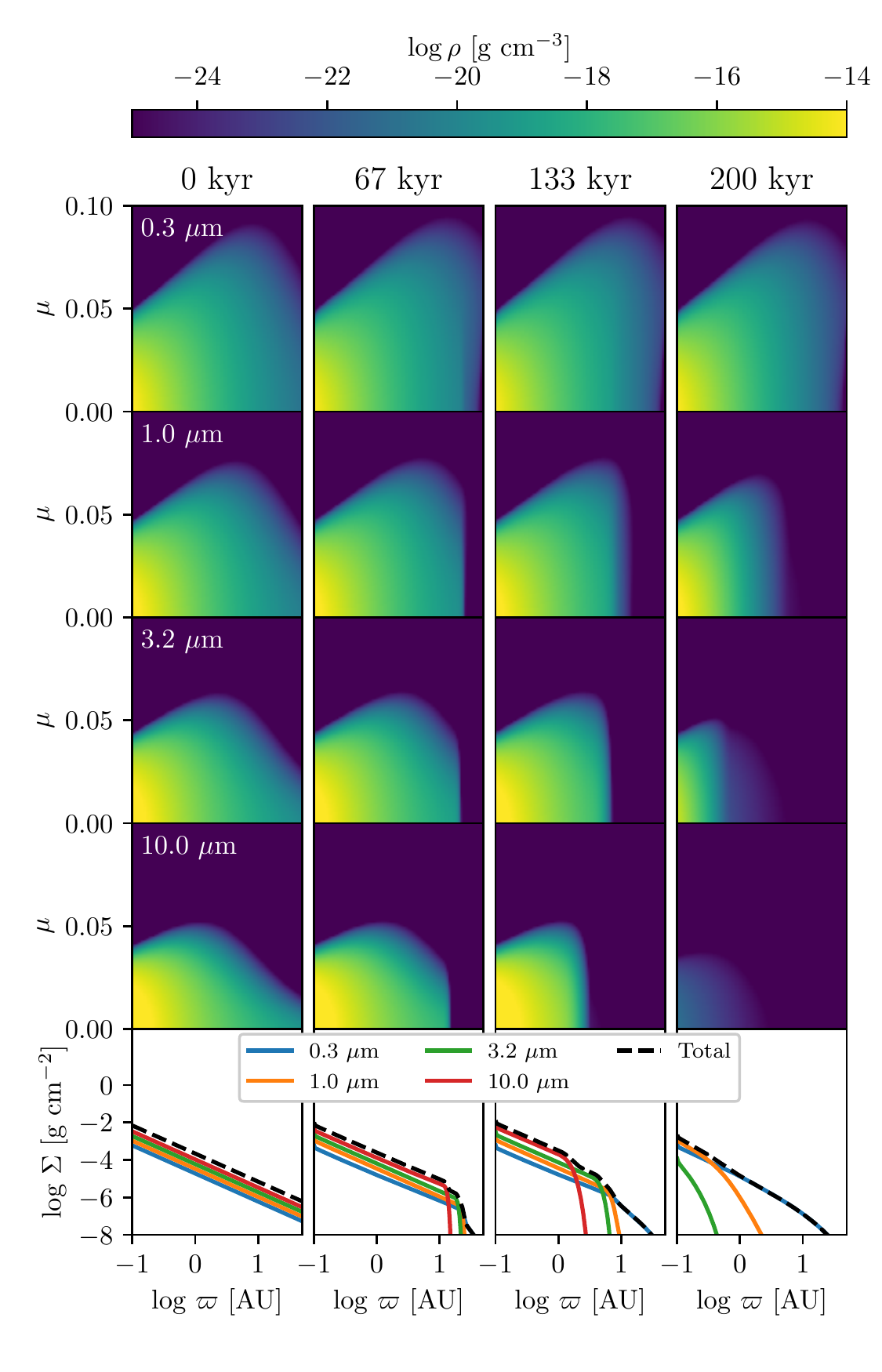}
    \caption{
    Same as \autoref{fig:eps2_fd5_slices} (case A2), but for the run with $\dot{M}=10^{-9}$ $M_\odot$ yr$^{-1}$ (case C1 in \autoref{tab:2d_params}); note the very different evolution in the two cases. An animated version of this figure is included in the Supplementary material (online).}
    \label{fig:c1_slices}
\end{figure}

\begin{figure}
    \centering
    \includegraphics[width=\columnwidth]{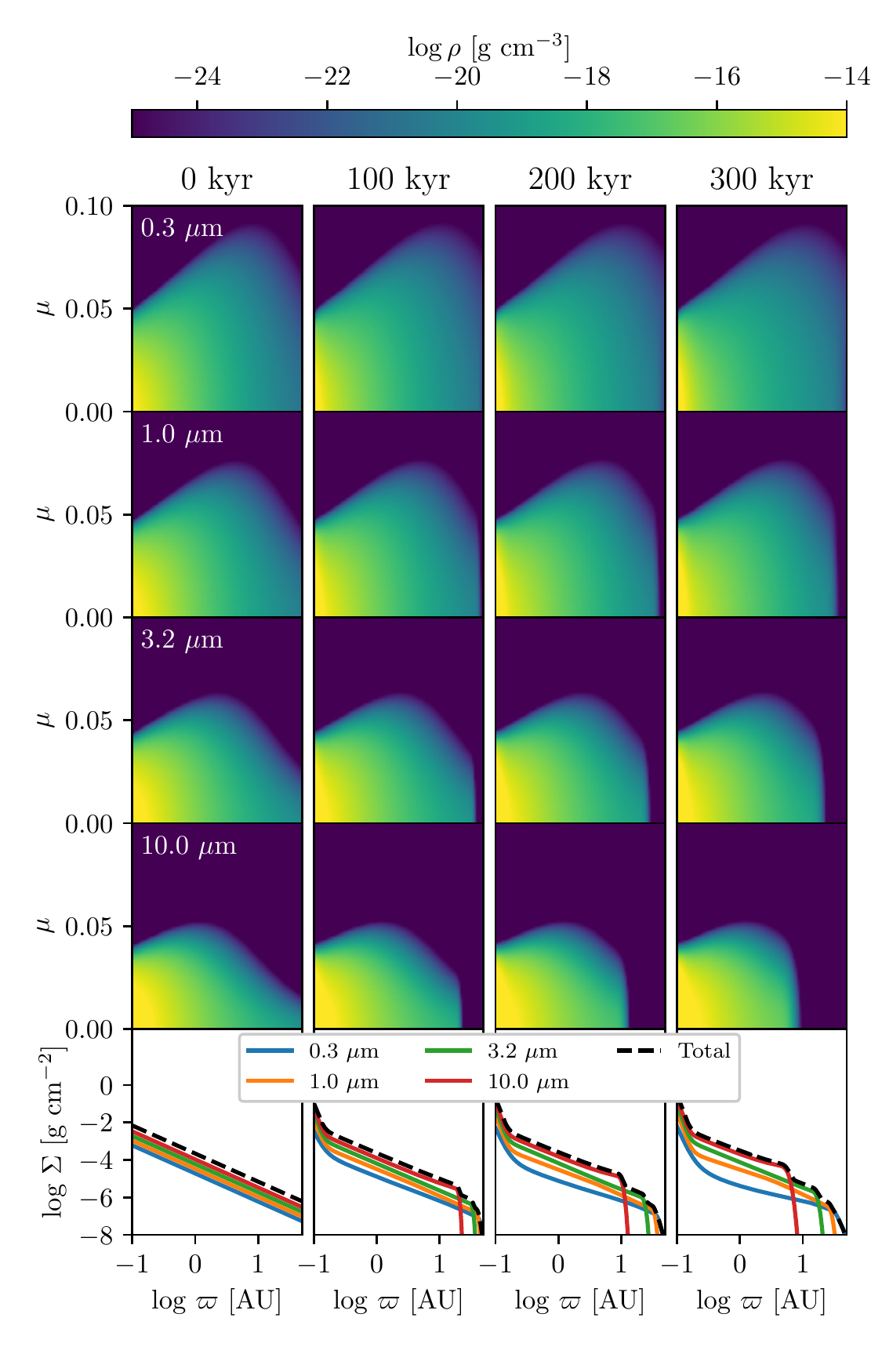}
    \caption{
    Same as \autoref{fig:eps2_fd5_slices} (case A2), but for the run with $\dot{M}=10^{-10}$ $M_\odot$ yr$^{-1}$ (case C2 in \autoref{tab:2d_params}); note how the outcome in this case is intermediate between that of case C1 ($\dot{M}=10^{-9}$ $M_\odot$ yr$^{-1}$) and A2 (no laminar accretion). An animated version of this figure is included in the Supplementary material (online).}
    \label{fig:c2_slices}
\end{figure}

\begin{figure}
    \centering
    \includegraphics[width=\columnwidth]{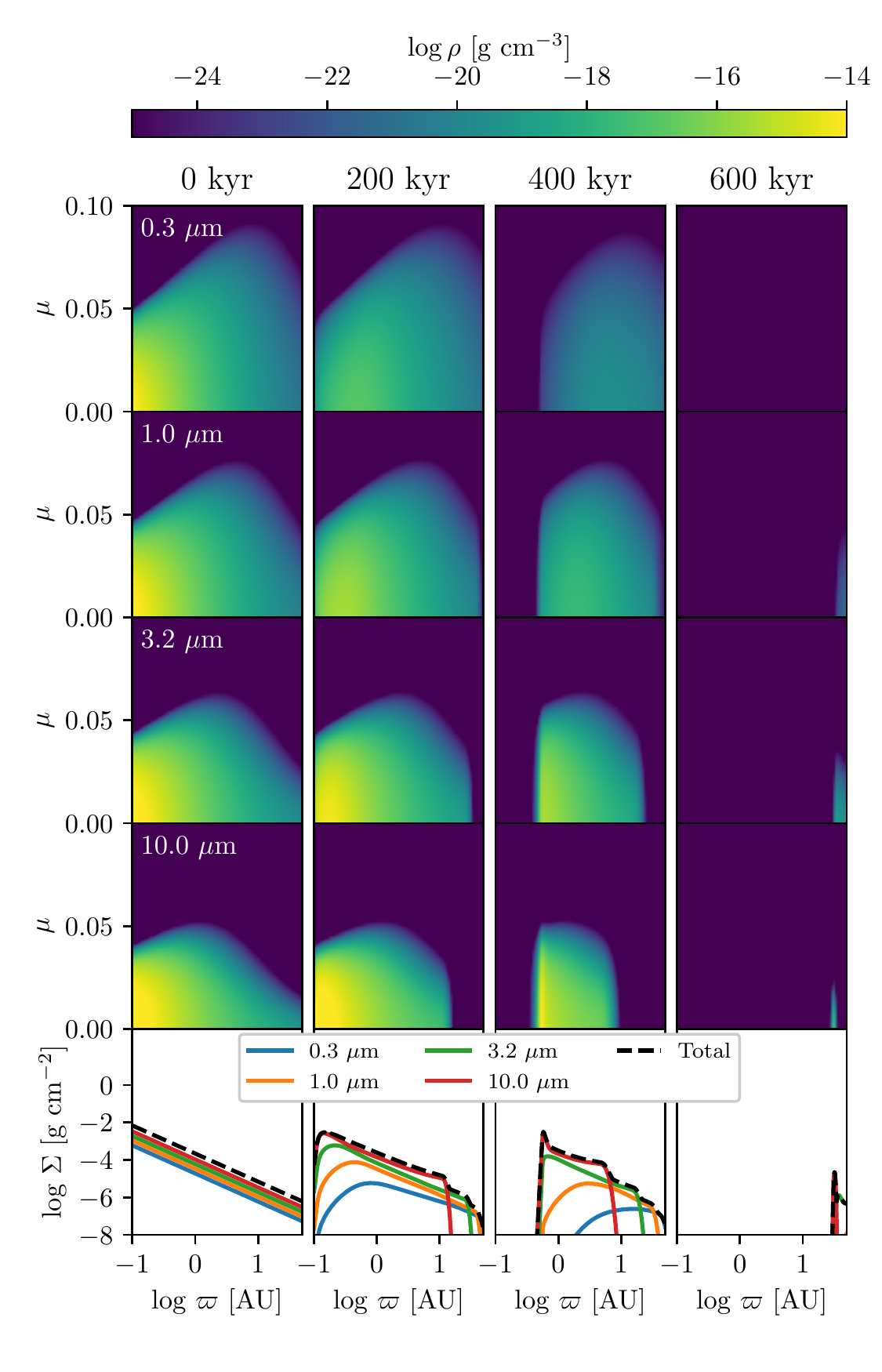}
    \caption{
    Same as \autoref{fig:eps2_fd5_slices} (case A2), but for the run with $\dot{M}=10^{-11}$ $M_\odot$ yr$^{-1}$ (case C3 in \autoref{tab:2d_params}); note that the two figures are nearly identical, indicating the minimal effect of laminar accretion at $10^{-11}$ $M_\odot$ yr$^{-1}$. An animated version of this figure is included in the Supplementary material (online).}
    \label{fig:c3_slices}
\end{figure}

\begin{figure}
    \centering
    \includegraphics[width=\columnwidth]{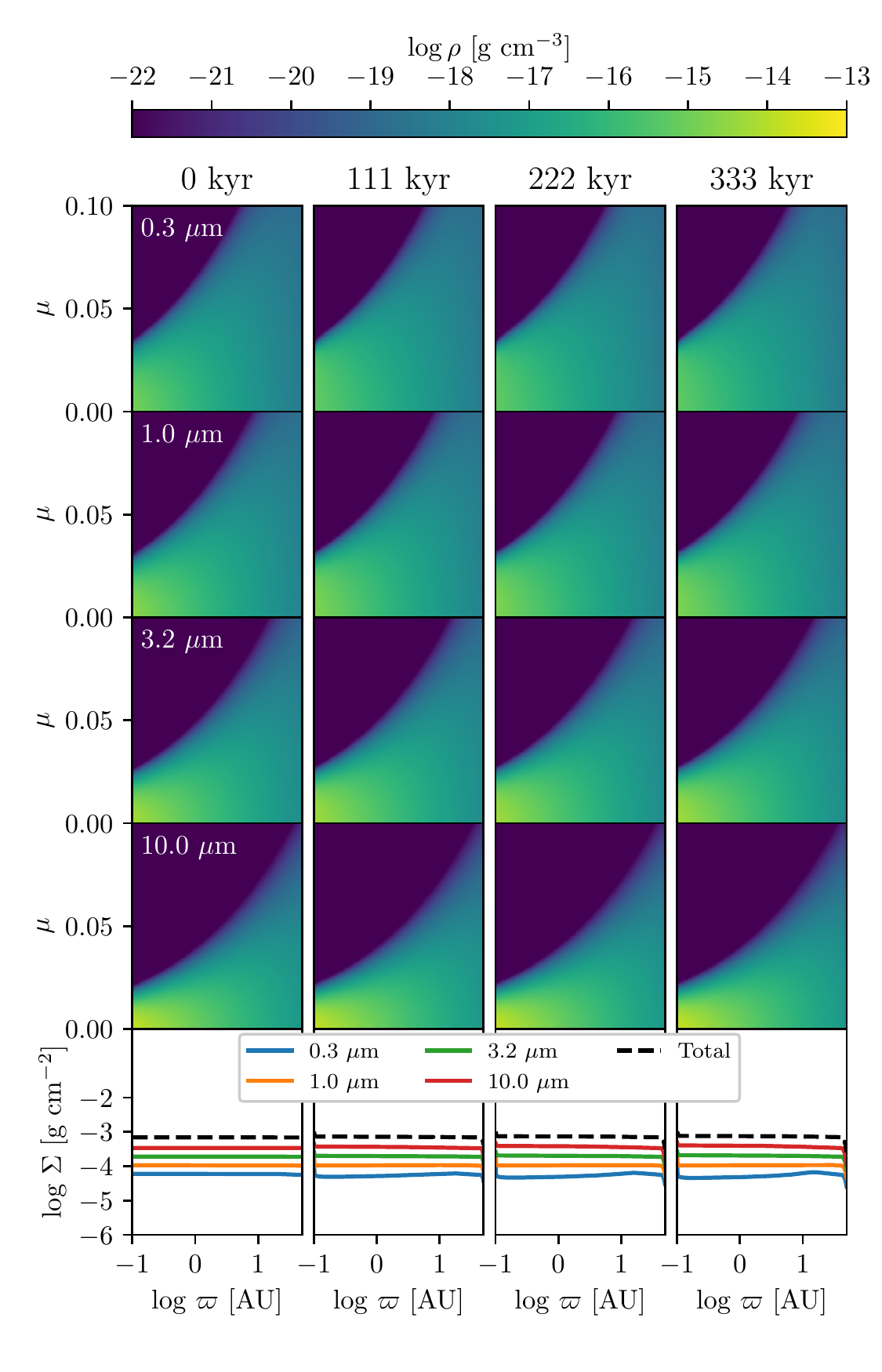}
    \caption{
    Same as \autoref{fig:eps25p0_fd4_slices} (case B2), but for the run with $\dot{M}=10^{-9}$ $M_\odot$ yr$^{-1}$ (case D1 in \autoref{tab:2d_params}); note that in this case the laminar inflow overwhelms the effects of radiation pressure, so the disc remains in steady state. An animated version of this figure is included in the Supplementary material (online).}
    \label{fig:d1_slices}
\end{figure}

\begin{figure}
    \centering
    \includegraphics[width=\columnwidth]{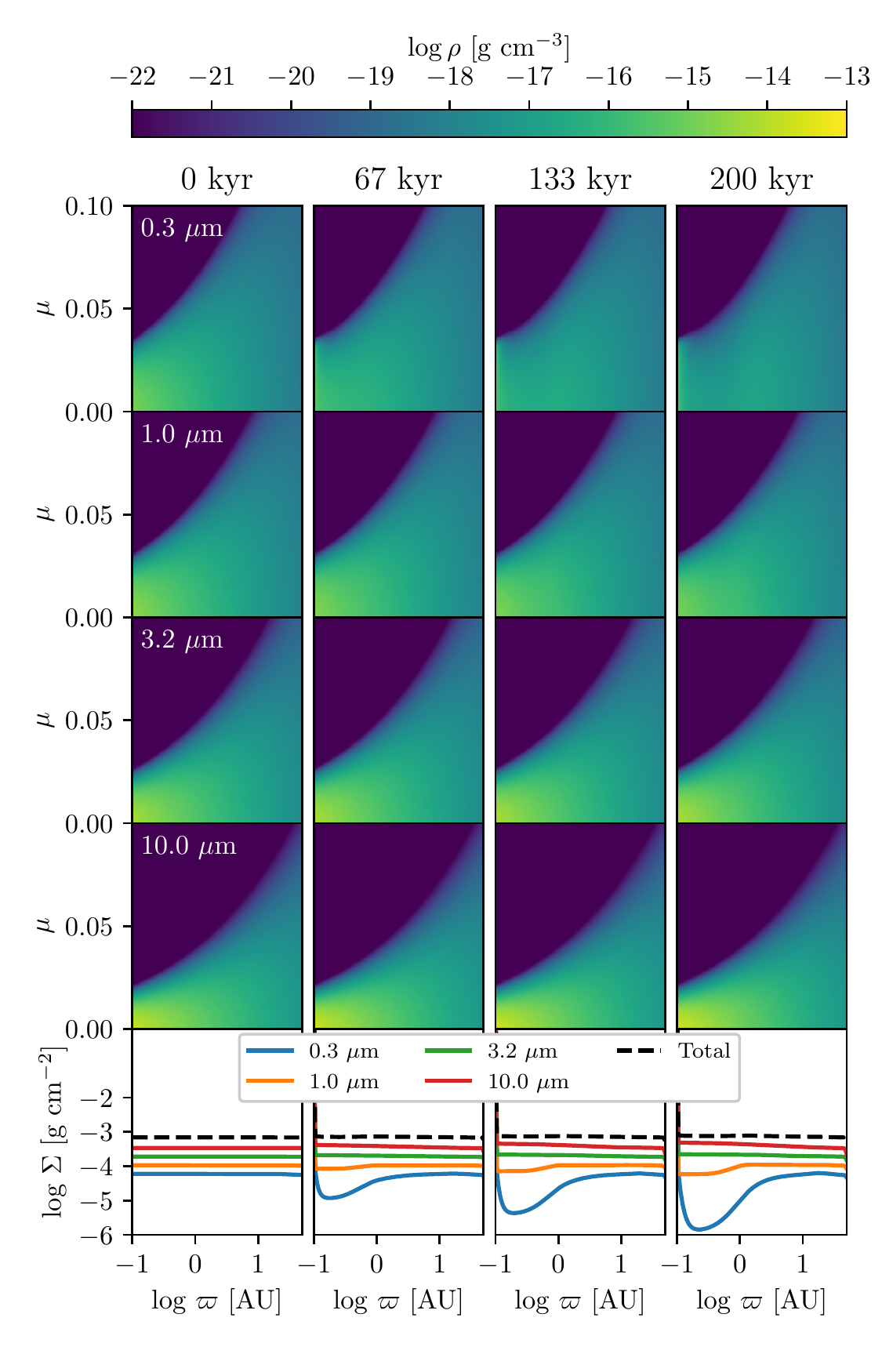}
    \caption{
    Same as \autoref{fig:eps25p0_fd4_slices} (case B2), but for the run with $\dot{M}=10^{-10}$ $M_\odot$ yr$^{-1}$ (case D2 in \autoref{tab:2d_params}); note how the outcome in this case is intermediate between that of case D1 ($\dot{M}=10^{-9}$ $M_\odot$ yr$^{-1}$) and B2 (no laminar accretion). An animated version of this figure is included in the Supplementary material (online).}
    \label{fig:d2_slices}
\end{figure}

\begin{figure}
    \centering
    \includegraphics[width=\columnwidth]{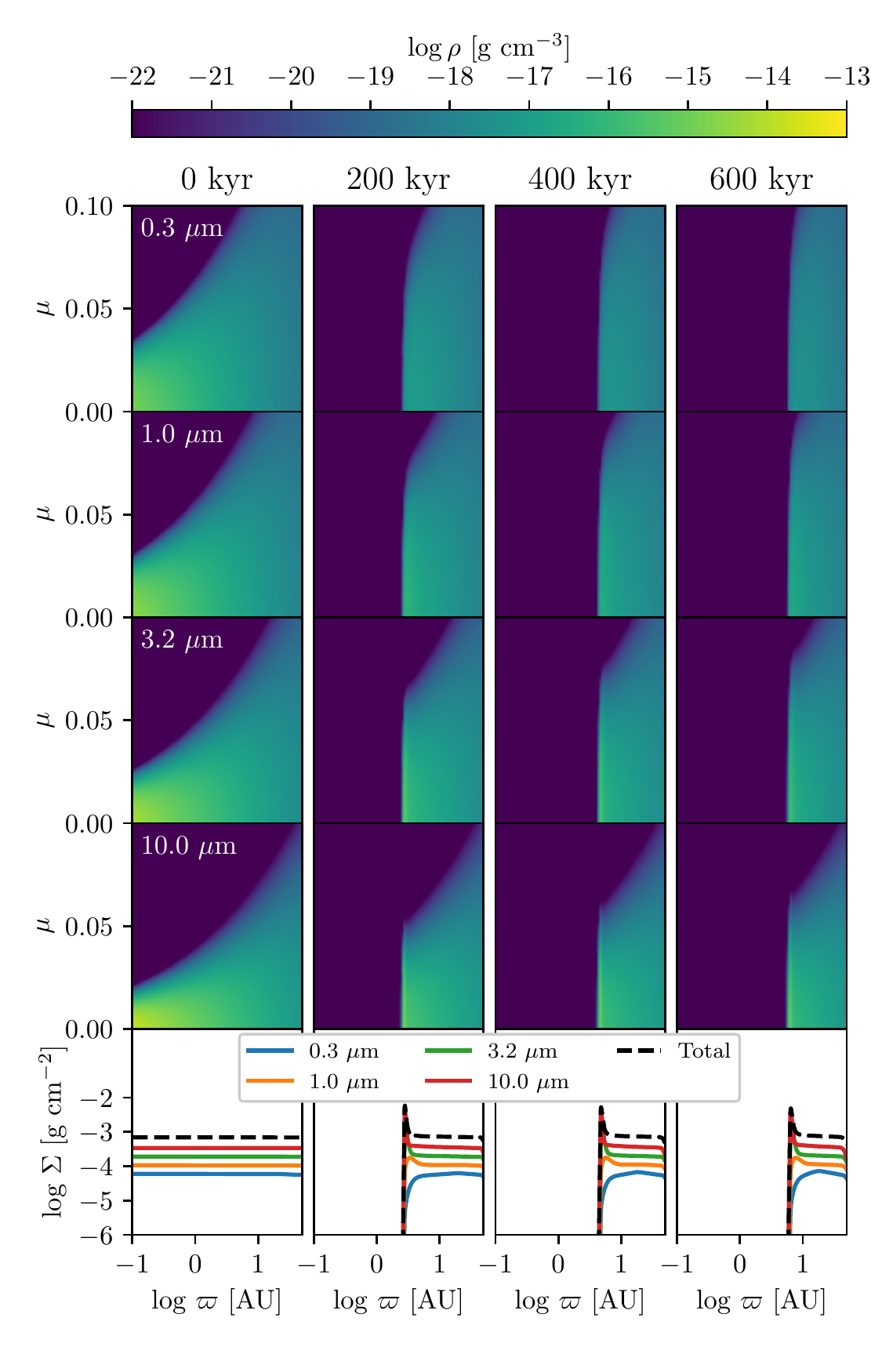}
    \caption{
    Same as \autoref{fig:eps25p0_fd4_slices}, but for the run with $\dot{M}=10^{-11}$ $M_\odot$ yr$^{-1}$ (case D3 in \autoref{tab:2d_params}); note that the two figures are nearly identical, indicating the minimal effect of laminar accretion at $10^{-11}$ $M_\odot$ yr$^{-1}$. An animated version of this figure is included in the Supplementary material (online).}
    \label{fig:d3_slices}
\end{figure}

Our series with laminar inflow, series C and D, use the same initial conditions as cases A2 and B2, respectively, but add laminar inflow at rates of $10^{-11} - 10^{-9}$ $M_\odot$ yr$^{-1}$. These cases therefore represent intermediate choices for initial disk properties, with inflow rates that range from the low to the moderate end of observationally-inferred values for transition discs. We show the results of series C, for which the initial disc is centrally-concentrated, in \autoref{fig:c1_slices} - \autoref{fig:c3_slices}, and the corresponding results for series D, with an initially-flat surface density distribution, in \autoref{fig:d1_slices} - \autoref{fig:d3_slices}.

Qualitatively, we see that cases C3 and D3, with accretion rates of $10^{-11}$ $M_\odot$ yr$^{-1}$, are almost identical to their analogs without laminar inflow, cases A3 and B3 -- for example compare \autoref{fig:eps2_fd5_slices} with \autoref{fig:c3_slices}, and \autoref{fig:eps25p0_fd4_slices} with \autoref{fig:d3_slices}. Such a low accretion rate has no significant effects on the system. By contrast, at a laminar accretion rate of $10^{-9}$ $M_\odot$ yr$^{-1}$, the differences are much larger. Comparing cases A2 (\autoref{fig:eps2_fd5_slices}) and C1 (\autoref{fig:c1_slices}), which have identical initial conditions, we see that accretion prevents the formation of a radiation-driven cavity, as happens in the non-accretion run. Instead, the flow pattern is that small grains circulate: near the midplane where density is high and radiation force is limited by shielding, drag forces win and grains move inward. Above the midplane, however, the drag force and shielding rapidly diminishes, and radiation pushes grains outward, whereupon they diffuse back toward the midplane. This circulation is most efficient for the smallest grains, which sit highest in the disc, and leads to a pattern whereby the outer disc becomes dominated by small grains because they are replenished more efficiently than larger grains. On the other hand, when compare cases D1 (\autoref{fig:d1_slices}) and B2 (\autoref{fig:eps25p0_fd4_slices}), we find that radiation pressure is ineffective at all radii, and the disc simply maintains a steady state, other than some minor depletion of the outer disc caused by accretion. The difference between cases C1 and D1 is driven by the initial surface density profile, which implies differing radial variation of the gas inflow velocity: $v_{g,\varpi} \propto 1/\varpi \Sigma_g$. For case C1 we have $p=-1.5$, so $v_{g,\varpi}\propto \varpi^{0.5}$, while for D1 we have $p=0$ and $v_{g,\varpi}\propto \varpi^{-1}$. Case D1 thus features significantly higher gas velocities at small radii, explaining why radiation pressure is ineffective in this case.

Our two cases with an inflow rate of $10^{-10}$ $M_\odot$ yr$^{-1}$, C2 and D2, and intermediate between the more and less rapidly accreting cases. Examining C2, \autoref{fig:c2_slices}, we see that radiation is able to flow the inflow at small radii, leading to the buildup of a region of enhanced surface density near the star, unlike in case C1. However, it is unable to push this ring back from the star, as happens in case C3. Similarly, in D2, \autoref{fig:d2_slices}, we see that radiation has little on the larger grains, except for creating a small density bump at the smallest radii, but efficiently repels grains in our smallest size bin from the star, leading to a central hole in small grains only. Thus the outcome in case D2 is similar to that in D1 for large grains (i.e., radiation has no effect and the disc remains steady), and similar to that in D3 for small grains (i.e., a hole opens). 

\section{Discussion}

\subsection{Astrophysical Implications}

First consider the results in the absence of a laminar accretion flow. In our simplified one dimensional case, we found that grains clear faster than they accrete for a dimensionless parameter $\chi\gtrsim 10$ (\autoref{eq:chi}). Considering now the 2D models of \autoref{sec:2D}, it is helpful to keep in mind that the accretion timescale $t_\text{acc}$ in physical units was $\sim 3$ Myr for our relatively low viscosity parameter $\alpha=10^{-4}$, so a more relevant criterion for whether radiative dust clearing is significant is arguably whether the clearing timescale $t_\text{clr}$ is on the same order of magnitude as the $\sim 0.1$ Myr timescale for the transitional disk phase as constrained by population studies \citep{Alexander14}. This timescale $t_\text{clr}$ has no dependence on $\alpha$ for high $\chi$, and is proportional to $\epsilon^{2} f_d$ (\autoref{eq:tclr}, noting that $\rho_g \propto \epsilon$). For our case with gas surface density power law index $p=-1.5$, in our most gas- and dust-rich case, case A1, we have $\epsilon^{2} (f_d/0.01) =10^{-4}$, while the values are $10^{-5}$ for case A2 and $10^{-6}$ for case A3. The simulations show that a transition to rapid dust clearing occurs around $\epsilon^{2} (f_d/0.01) \sim 10^{-5}$. For our $p=0$ cases, models B1 - B3, we also find a transition to efficient clearing around $\epsilon^{2} (f_d/0.01) \sim 10^{-5}$, corresponding to case B2 (c.f.~\autoref{tab:2d_params}). Thus our simulations, together with our analytical calculation of timescales and their dependence on disc dust and gas properties, support the general hypothesis that radiative dust clearing is a significant process in any non-accreting disc satisfying $\epsilon^{2} (f_d/0.01) \lesssim 10^{-5}$ in the inner $\sim$1\,AU from the star.

If we now add a laminar accretion flow to this picture, this adds an additional condition. Since we expect the ratio of drag force to radiation force to scale as roughly $\dot{M}/L$, we phrase this condition in terms of the value of this parameter. When we add such a flow to cases that lie on the boundary of dust clearing, A2 and B2, we find that an accretion-luminosity ratio of $\dot{M}/L = 10^{-11}$ $M_\odot$ yr$^{-1}$ $L_\odot^{-1}$ leaves the results qualitatively unchanged. An accretion rate two orders of magnitude larger suppresses or completely negates the effects of radiation pressure, depending on the disc density profile, while an accretion rate of $\dot{M}/L = 10^{-10}$ $M_\odot$ yr$^{-1}$ $L_\odot^{-1}$ produces an intermediate result, whereby radiation has strong effects on smaller grains, but weaker effects on larger ones. In this regime radiation pressure is relatively unimportant in terms of the total grain mass budget, but is likely to be very important in terms of observed properties: small grains have a steep negative opacity power law ($\kappa \sim \lambda^{-1}$), and thus depleting sub-micron grains by a factor of $\sim 10-100$, as happens in runs C2 (\autoref{fig:c2_slices}) and D2 (\autoref{fig:d2_slices}), will likely reduce the influence of warm dust on the SED, decreasing the $\sim$2--5\,$\mu$m disc flux significantly. This is an obvious topic for future work including multi-grain species Monte-Carlo radiative transfer. Thus a rough summary of our findings is that radiative clearing is a significant effect, at least for sub-micron grains and the parts of the SED that are most sensitive to them, if the ratio of accretion rate to luminosity is $\lesssim 10^{-10}$ $M_\odot$ yr$^{-1}$ $L_\odot^{-1}$, but that a wide range of outcomes are possible near the boundary, depending on details such as the disc density profile. In all cases where radiation has any effects, however, we find strong sorting of grains by size.

Finally, we note that, in the critical inner $\sim 10$ AU, gas power law densities have not been carefully measured for Class II objects, although it appears possible to do so in the coming years with ALMA \citep{Miotello18}. For transitional discs, gas density models that combine spectra with partially resolved observations in CO isotopologs result in only moderately satisfactory fits \citep{van-der-Marel16a}, but clearly indicate very significant depletion of CO in the inner $\sim 10$ AU. This CO depletion is further supported by simultaneous modelling of spectra and spectro-astrometry \citep{Pontoppidan08}, where the complete lack of CO gas at high velocities is strong evidence of cleared CO within $\sim 5$ AU of SR~21 in particular. Typical values of $\epsilon$ around $10^{-2.5}$ and power law indices between 0 and $-1.5$, i.e., precisely the range spanned by our simulations, are consistent with those papers. Small dust grains are also severely depleted in these discs, with depletions in the very inner disc between $10^2$ and $10^6$; indeed, spectral energy distribution modelling is consistent with there being no dust at all at moderate radii ($\sim 10$ AU; \citealt{van-der-Marel16a}). The mechanisms we have explored in this paper provide a natural explanation for these results, since we show that the combination of radiative acceleration, gas drag and turbulent viscosity could start from the small grain dust distribution of a Class II T~Tauri star and produce an inner hole largely cleared of small grains, so long as some grain growth ($f_d \sim 10^{-4}$) and accretion-based gas clearing ($\epsilon \sim 10^{-2.5}$) occurs during the Class II phase.


\subsection{Model Limitations}

We end this discussion by pointing out some of the limitations of the models we have explored thus far, which point to the directions required in future work. First, our calculation is limited to grains whose interaction with radiation can be approximated by geometric optics. Although this is an excellent starting point, since it applies to almost all grains larger than a few tenths of a micron, it does not represent the situation in the most evolved of discs, where for example there is evidence for very small grains \citep[Oph~IRS~48 and HD~169142,][]{Birchall19} or unusually bright scattering indicative of unusual optical properties \citep[LkCa~15,][]{thalmann_resolving_2016}. Indeed, the mechanism proposed in this paper provides a natural explanation for clearing out grains with a ``normal'' optical properties, and thus high ratio of radiative to gravitational acceleration $\beta$, leaving unusually low $\beta$ grains behind. 

A second limitation of our models is that we considered a static background disc, rather than one whose structure is self-consistently generated as a result of viscous accretion and similar processes that shape the gas distribution. This makes it difficult to directly relate our model parameters to the stellar accretion rate, which is a key measurable parameter of real star-disk systems. We chose not to evolve a model where the gas was in a viscous steady state because although turbulent diffusion is certainly a key driver of the evolution of the dust density distribution, turbulent viscosity \citep{Shakura73} is not the leading candidate for driving gas accretion in cool, evolved disks approaching or in the transitional phase \citep{Bai13,Turner14}. Ideally, the work in this paper could be coupled with a plausible gas accretion mechanism. We note that gas clearing may be coupled with dust clearing, as the dust density distribution directly feeds back to the true gas scale height and dynamics -- another physical mechanism beyond the scope of this paper.

A third limitation worth mentioning, which is related to the previous one, is that real disks will also likely have inflow velocities and turbulence properties that vary somewhat with height, due to either disk winds, partial MRI active layers, or even hydrodynamic effects such as the vertical shear instability \citep{Nelson:2013,Bai2017,Mohanty:2018,Gressel20}. Thus it is possible that a high accretion rate might be maintained through the midplane, while radiatively driven dust evolution substantially alters the dust distribution in the disk atmosphere, as in our non-accreting simulations. Such shear, in the absence of a strong stabilizing temperature inversion, could produce circulation and mixing between layers. A full accounting of the impact of highly vertically dependent accretion is beyond the scope of this work.

\section{Conclusions}

We have shown in this paper that as grain growth and accretion processes naturally clear a protoplanetary disc, there is a transition point where the combination of radiation pressure and gas drag can rapidly remove small ($\sim$micron) sized grains from the inner disc, leaving a transitional disc structure behind. The physical mechanism for this clearing is a reduction in the effective gravity of small grains, resulting in a smaller orbital velocity so that gas drag can move the grains outwards. Using 2D simulations in which we simultaneously include radiative forces, turbulent diffusion of dust by gas, and inward flow of dust due to gas accretion and radial pressure gradients, we show that the disc clearing is not simply a surface effect, and can affect the entire small grain dust disc structure. This process had not been studied in depth before, because the mechanism alone is not effective for a minimum mass solar nebula, and requires the disc to already have evolved significantly and to have substantially reduced its gas accretion rate compared to that during earlier evolutionary phases. Conversely, however, once these evolutionary processes drive the accretion rate and the gas and dust density low enough, radiative clearing becomes both unavoidable and rapid.

Our proposed physical clearing process has a number of appealing features. It does not invoke planet formation directly, and can take place even in a disc that does not form planets. It leaves behind a structure that is consistent with current transitional disc observations. Notably, this process clears dust and not gas, so is consistent with transitional discs still having moderately large gas discs while having an inner cavity that is almost completely devoid of dust.

The primary limitation of our work thus far is that, while we have considered a range of dust grain sizes, we have thus far limited our calculations to grains whose interaction with starlight can be described by geometric optics. Future work will involve relaxing this assumption allowing us to consider not just grains smaller than $\sim 0.1$ $\mu$m that are too small for geometric optics, but also grains with differing radiative properties, for example high degrees of scattering asymmetry. This will also enable radiative transfer models to see how effectively the dust structures produced by this model can reproduce real observations. 

\section*{Acknowledgements}

MRK and MJI acknowledge support from the Australian Research Council (MRK: FT180100375 and DP190101258; MJI: FT130100235 and DP170102233). MRK acknowledges support from an Alexander von Humboldt Research Award. KMK acknowledges support from the Mt. Stromlo Distinguished Visitor program. This research was undertaken with the assistance of resources from the National Computational Infrastructure (NCI Australia), an NCRIS enabled capability supported by the Australian Government. The software used for this project makes use of Astropy,\footnote{http://www.astropy.org} a community-developed core Python package for Astronomy \citep{Astropy-Collaboration13a, Astropy-Collaboration18a}.

\section*{Data availability}

The code we use to carry out all of the simulations presented in this paper is available from \url{https://bitbucket.org/krumholz/dustevol/} under an open source license. Simulations were run on the National Computational Infrastructure (NCI Australia), and full simulation results are available upon reasonable request from the corresponding author, but are not included in the public repository due to their size.




\bibliographystyle{mnras}
\bibliography{KMK_MAIN,MRK_refs,MJI_refs}



\appendix

\section{Numerical Method for 1D Systems}
\label{app:numerics_1d}

Here we describe the numerical method we use to solve the simplified 1D system, \autoref{eq:advec_diff_1d_nondim}. To avoid clutter in this appendix we drop the primes on all the terms in this equation, but all the quantities listed are the non-dimensonalised ones. 

\subsection{Spatial discretisation, initial conditions, and boundary conditions}

We use a uniform grid with constant cell size $\Delta x$, with the left edge of cell $0$ at $x=0$ and the right edge of cell $N-1$ at $x = x_{\rm max}$; we use $x_{i-1/2}$ and $x_{i+1/2}$ to denote the positions of the left and right edges of cell $i$. We use a finite volume discretisation on this grid; integrating \autoref{eq:advec_diff_1d_nondim} over cell $i$, we have
\begin{eqnarray}
\frac{\partial}{\partial t} \rho_{d,i} & = &
\Delta x^{-1} \left\{ \chi 
\left[\left(\rho_d e^{-\tau}\right)_{i+1/2}
- \left(\rho_d e^{-\tau}\right)_{i-1/2}
\right]
\vphantom{\left(\frac{\partial\rho_d}{\partial x}\right)_{i+1/2}}
\right.
\nonumber \\
& &
\left.
\qquad {}
- \left(\frac{\partial\rho_d}{\partial x}\right)_{i+1/2} + 
\left(\frac{\partial\rho_d}{\partial x}\right)_{i-1/2}
\right\},
\label{eq:advec_diff_1d_disc}
\end{eqnarray}
where $\rho_{d,i} = \Delta x^{-1} \int_{x_{i-1/2}}^{x_{i+1/2}} \rho_d \, dx$ is the average of $\rho_d$ over cell $i$, and the subscripts $i+1/2$ and $i-1/2$ indicate that a particular quantity is to be evaluated the corresponding cell edge.

We initialise the simulation with $\rho_{d,i} = 1$ in every cell, and adopt boundary conditions whereby both the advective and diffusive fluxes out of the domain are set to zero. To ensure that these choices do not affect the result, we always choose the size of our domain large enough so that the right edge of the domain is well beyond the dust wave, and thus the flux through cells near it is negligibly small in any event.

As the simulation evolves, and the dust front moves to larger $x$, an increasingly large fraction of the computational domain becomes filled with cells for which $\rho_{d,i} \approx 0$. To avoid expending CPU cycles needlessly updating these nearly-empty cells, at the end of each time step (see next section) we shift our grid to the left, removing the leftmost cells within which $\rho_{d,i} < 10^{-6}$. We keep the number of cells constant by adding an equal number of new cells on the right hand side of the domain, all initialised to $\rho_{d,i} = 1$; since our domains extend well past the edge of the dust wave at all times, the existing cells adjacent to those being added also have $\rho_{d,i}\approx 1$, and thus the newly-added cells blend smoothly with the existing ones.

\subsection{Time discretisation and time-stepping strategy}
\label{ssec:time_stepping}

We advance the calculation in time using an implicit-explicit update step with \citet{Strang68a} splitting between the advective terms, which we handle explicitly, and the diffusion terms, which we handle implicitly. To advance the calculation from time $t_n$ to time $t_{n+1} = t_{n} + \Delta t$, starting from the dust densities $\rho_{d,i}^{(n)}$ in every cell at time $n$, we carry out the following steps:
\begin{enumerate}
\item Advance the diffusion subsystem (the $\partial \rho_d/\partial x$ terms in \autoref{eq:advec_diff_1d_disc}) for a time $\Delta t/2$ using an implicit method that is second-order accurate in space and time (\autoref{sssec:diff_subsystem_1d}). We denote the state after this step as $\rho_{d,i}^{(*)}$.
\item Advance the advection subsystem (the $\rho_d e^{-\tau}$ terms in \autoref{eq:advec_diff_1d_disc}), starting from state $\rho_{d,i}^{(*)}$, for a time $\Delta t$ using an explicit method that is second-order accurate in time and third-order accurate in space (\autoref{sssec:adv_subsystem_1d}). We denote the state that results from this procedure $\rho_{d,i}^{(\dagger)}$.
\item Starting from state $\rho_{d,i}^{(\dagger)}$, carry out a final diffusion advance, using the same method as in step 1, through time $\Delta t/2$. This yields the final state $\rho_{d,i}^{(n+1)}$ at the new time.
\end{enumerate}
The overall scheme is second-order accurate in time. We set the time step based on a Courant-Friedrichs-Lewy (CFL) condition as applied to the advection step, which we describe in \autoref{sssec:adv_subsystem_1d}. This condition guarantees positivity during the first advection update, but is not sufficient to guarantee positivity through the full Strang-split time step. Thus on occasion our update scheme yields a negative density. If this occurs, we simply reduce the time step by factors of 2 and retry until the step succeeds.

\subsection{Subsystems}
\label{ssec:subsystems_1d}

Here we describe our procedures for advancing the advection and diffusion subsystems.

\subsubsection{Diffusion}
\label{sssec:diff_subsystem_1d}

We evaluate the diffusion terms in \autoref{eq:advec_diff_1d_disc} using second-order accurate centered differences,
\begin{equation}
\left(\frac{\partial \rho_d}{\partial x}\right)_{i+1/2} = \frac{\rho_{d,i+1}-\rho_{d,i}}{\Delta x},
\end{equation}
and similarly for cell edge $i-1/2$. We discretise the diffusion subsystem in time using a second-order accurate Crank-Nicolson method. Defining $\Theta = 1/2$ as the time centring parameter, for time step $\Delta t$ we obtain the usual second-order Cartesian diffusion discretisation:
\begin{eqnarray}
\lefteqn{
\rho_{d,i}^{(n+1)} = \rho_{d,i}^{(n)}
- \left(\frac{\Delta t}{\Delta x^2}\right) \cdot {}
}
\nonumber \\
& & 
\left[
\left(1-\Theta\right)
\left(\rho_{d,i+1} - 2 \rho_{d,i} + \rho_{d,i-1}\right)^{(n)}
\right.
\nonumber \\
& &
\left.
{}
+
\Theta
\left(\rho_{d,i+1} - 2 \rho_{d,i} + \rho_{d,i-1}\right)^{(n+1)}
\right],
\end{eqnarray}
where the superscript $(n)$ indicates the state at the start of the update and $(n+1)$ indicates the state after the diffusion update. This equation can be rewritten as a linear system
\begin{equation}
\label{eq:mat_diff_1d}
\matA \vecrho_{d}^{(n+1)} = \vecrho^{(n)} + \Delta t \dot{\vecrho}^{(n)}_{{\rm diff}},
\end{equation}
where $\vecrho_d$ is an $N$-element vector containing the densities in all cells,
\begin{equation}
\dot{\vecrho}_{\rm diff}^{(n)} = 
\left(1 - \Theta\right)
\left(\frac{\rho_{d,i+1} - 2 \rho_{d,i} + \rho_{d,i-1}}{\Delta x^2}\right)^{(n)}
\end{equation}
is the rate of change of $\vecrho_d$ due to diffusion evaluated at the old time, and the $\matA$ is an $N\times N$ tridiagonal matrix whose elements are
\begin{equation}
M_{ij} = \left(1 + 2 \Theta \frac{\Delta t}{\Delta x^2}\right) \delta_{ij} - \Theta \frac{\Delta t}{\Delta x^2} \left(\delta_{i,j+1} + \delta_{i,j-1}\right).
\end{equation}
We solve \autoref{eq:mat_diff_1d} using the standard Thomas algorithm for tridiagonal matrices.

\subsubsection{Advection}
\label{sssec:adv_subsystem_1d}

Since the velocities are non-local functions of the density coupled via the radiation field, we cannot obtain second-order accuracy in time using standard predictor-corrector methods. Instead, following \citet{Skinner18a}, we use a \citet{Shu88a} TVD time discretisation. Given a starting state $\rho_{d,i}^{(n)}$, we advance the calculation from $t_{n}$ to $t_{n+1} = t_n + \Delta t$ via
\begin{eqnarray}
\rho_{d,i}^{(*)} & = & \rho_{d,i}^{(n)} + \Delta t \left(\dot{\rho}_{\rm adv}\right)_{d,i}^{(n)} \\
\rho_{d,i}^{(n+1)} & = & \frac{1}{2}\rho_{d,i}^{(n)} + \frac{1}{2}\rho_{d,i}^{(*)} + \frac{1}{2}\Delta t \left(\dot{\rho}_{\rm adv}\right)_{i}^{(*)},
\end{eqnarray}
where
\begin{equation}
\left(\dot{\rho}_{\rm adv}\right)_i^{(n)} = \frac{\chi}{\Delta x} \left[\left(\rho_d e^{-\tau}\right)_{i+1/2} - \left(\rho_d e^{-\tau}\right)_{i-1/2}\right]^{(n)}
\end{equation}
is the rate of change in $\rho_{d,i}$ evaluated using the density field $\rho_{i}^{(n)}$, and similarly for $(\dot{\rho}_{\rm adv})_{i}^{(*)}$.

We evaluate these terms in three steps. First, for the initial density field $\rho_{d,i}^{(n)}$ or $\rho_{d,i}^{(*)}$ we compute a piecewise-parabolic method (PPM) reconstruction of the density field \citep{Colella84a}. Specifically, we approximate the dust density in cell $i$ with a parabolic function
\begin{equation}
\rho_{d}(x) = c_{0,i} + c_{1,i} x + c_{2,i} x^2,
\end{equation}
where the reconstruction has the property that $\Delta x^{-1} \int_{x_{i-1/2}}^{x_{i+1/2}} \rho_d(x)\, dx = \rho_{d,i}$, and the function $\rho_d(x)$ is monotonic for $x \in (x_{i-1/2},x_{i+1/2})$. The reconstruction coefficients $c_0$, $c_1$, and $c_2$ are functions of $\rho_{d,i-1}$, $\rho_{d,i}$, and $\rho_{d,i+1}$. Second, we calculate the optical depths to cell edges as 
\begin{equation}
\tau_{i+1/2} = \sum_{j=0}^i \rho_{d,i}.
\end{equation}
From the optical depths we can derive the velocities $v_{i+1/2} = \chi e^{-\tau_{i+1/2}}$ at each cell face.
Third, we evaluate the mass per unit area transported across cell face $i+1/2$ over time $\Delta t$, which is $\Delta t \left(\rho_d e^{-\tau}\right)_{i+1/2}$, by evaluating the amount of mass contained in the region between the cell edge and the position of a contact discontinuity propagating at speed $v_{i+1/2}$ away from that cell edge:
\begin{equation}
\Delta t \left(\rho_d e^{-\tau}\right)_{i+1/2} = \int_{x_{i+1/2} - v_{i+1/2}\Delta t}^{x_{i+1/2}} \rho_d(x) \, dx.
\end{equation}
This procedure allows us to evaluate $\left(\dot{\rho}_{\rm adv}\right)_i$ in every cell.

We set the overall time step as
\begin{equation}
    \Delta t = C \min\left(\frac{r_{i}-r_{i-1}}{|v_{i+1/2}|}\right),
    \label{eq:timestep}
\end{equation}
where the minimum is over all cells $i$, and $C$ is the CFL number. The scheme is stable for $C < 0.5$.

\section{Numerical Method for 2D Systems}
\label{app:numerics}

Here we describe the full details of the method we use to solve the advection-diffusion equation, \autoref{eq:advec_diff}. Our strategy here is a 2D generalisation of that described in \aref{app:numerics_1d}.

\subsection{Computational grid and discretisation}

We adopt a 2D spherical polar grid with $(r,\theta)$ as our basic coordinates, but we will specify our grid in terms of the radial and angular variables $x$ and $\mu$, where $x = \log r/r_{\rm min}$, $\mu = \sin\theta$, and $r_{\rm min}$ is the inner edge of the computational domain. The centre of cell $ij$ is located at coordinates $x_i$ and $\mu_j$, and its upper right corner is located at $x_{i+1/2}$ and $\mu_{j+1/2}$. In the radial direction the grid starts at $r_{-1/2} = r_{\rm min}$ and ends at $r_{N_r-1/2} = r_{\rm max}$, and in the azimuthal direction it extends from $\mu_{-1/2} = 0$ to $\mu_{N_\theta-1/2} = \mu_{\rm max}$; the grid is $N_r \times N_\theta$ cells in size, and is uniformly-spaced so that the sizes of cells are $\Delta x = (x_{N_r-1/2}-x_{-1/2})/N_r$ and $\Delta \mu = \mu_{N_\theta-1/2}/N_\theta$. For this grid, cell volumes and radial and angular cell face areas are
\begin{eqnarray}
V_{ij} & = & \frac{2}{3}\pi \left(r_{i+1/2}^3-r_{i-1/2}^3\right) \Delta \mu \\
A_{i+1/2,j} & = & 2\pi r_{i+1/2}^2 \Delta\mu \\
A_{i,j+1/2} & = & \pi \left(r_{i+1/2}^2 - r_{i-1/2}^2\right),
\end{eqnarray}
respectively.

We also discretize in bins of grain size, by defining a logarithmically-spaced set of grain size bins. Specifically, we use $N_a$ grain size bins, with $a_{-1/2} = a_{\rm min}$ representing the smallest size grains in the smallest bin, and $a_{N_a-1/2}$ the largest grains in the largest bin, $\Delta \log a = \log(a_{k+1}/a_k) = \log(a_{N_a-1/2}/a_{-1/2})/N_a$ constant, and $a_k = \sqrt{a_{k-1/2} a_{k+1/2}}$ representing the mean size of grains in the $k$th size bin. We let $\rho_{d,k}$ represent the mean density of grains in the size range from $a_{k-1/2}$ to $a_{k+1/2}$.

We adopt a finite-volume spatial discretization strategy. Integrating \autoref{eq:advec_diff} over the volume of cell $ij$, and making use of the divergence theorem, we have
\begin{eqnarray}
\lefteqn{
\frac{\partial}{\partial t} \rho_{d,ijk} = -V_{ij}^{-1} \cdot {}
}
\nonumber \\
& &
\left[
F_{{\rm adv},i+1/2,j,k} A_{i+1/2,j} - F_{{\rm adv},i-1/2,j,k} A_{i-1/2,j} + {}
\right.
\nonumber \\
& & \;\;
F_{{\rm adv},i,j+1/2,k} A_{i,j+1/2} - F_{{\rm adv},i,j-1/2,k} A_{i,j-1/2} + {}
\nonumber \\
& & \;\;
F_{{\rm diff},i+1/2,j,k} A_{i+1/2,j} - F_{{\rm diff},i-1/2,j,k} A_{i-1/2,j} + {}
\nonumber \\
& & \;\;
\left.
F_{{\rm diff},i,j+1/2,k} A_{i,j+1/2} - F_{{\rm diff},i,j-1/2,k} A_{i,j-1/2}
\right],
\label{eq:advec_diff_discrete_a}
\end{eqnarray}
where $F_{\rm adv}$ and $F_{\rm diff}$ are the advective and diffusive fluxes at the cell faces, defined by
\begin{eqnarray}
F_{{\rm adv},i\pm 1/2,j,k} & = & \int_{A_{i\pm 1/2,j}} \rho_d v_{d,r} \, dA\\
F_{{\rm adv},i,j\pm 1/2,k} & = & \int_{A_{i,j\pm 1/2}} \rho_d v_{d,\theta} \, dA\\
F_{{\rm diff},i\pm 1/2,j,k} & = & -\int_{A_{i\pm 1/2,j}} D_{d,a} \rho_g \frac{df_d}{dr}\, dA \\
F_{{\rm diff},i,j\pm 1/2,k} & = & -\int_{A_{i,j\pm 1/2}} D_{d,a} \rho_g  \frac{1}{r}\frac{df_d}{d\theta}\, dA
\end{eqnarray}
Here, for each size bin $k$, $\rho_{d,ijk}$ is the mean dust density in cell $ij$, $v_{d,r}$ and $v_{d,\theta}$ are the $r$ and $\theta$ velocities, and $D_{d,k}$ is the diffusion coefficient evaluated for grains of size $a_k$. Note that \autoref{eq:advec_diff_discrete_a} is exact. We defer discussion of how we evaluate the fluxes to \autoref{ssec:subsystems}.

We discretise the equations in time and advance using the same approach as in the 1D case, as described in \aref{ssec:time_stepping}. Specifically, we break the problem into advective and diffusive subsystems, and use \citet{Strang68a} splitting to advance them alternately while retaining second-order accuracy in time.

\subsection{Subsystems}
\label{ssec:subsystems}

Here we describe our procedures for advancing the advection and diffusion subsystems.

\subsubsection{Diffusion}
\label{sssec:diff_subsystem}

The diffusion subsystem of \autoref{eq:advec_diff_discrete_a} is
\begin{eqnarray}
\lefteqn{
\frac{\partial}{\partial t} \rho_{d,ijk} = -V_{ij}^{-1} \cdot {}
}
\nonumber \\
& &
\left[
F_{{\rm diff},i+1/2,j,k} A_{i+1/2,j} - F_{{\rm diff},i-1/2,j,k} A_{i-1/2,j} + {}
\right.
\nonumber \\
& & \;\;
\left.
F_{{\rm diff},i,j+1/2,k} A_{i,j+1/2} - F_{{\rm diff},i,j-1/2,k} A_{i,j-1/2}
\right].
\label{eq:diff_discrete_a}
\end{eqnarray}
We evaluate the diffusive fluxes using second-order accurate centred finite differences:
\begin{eqnarray}
F_{{\rm diff},i\pm 1/2,j,k} & = & -D_{d,i\pm 1/2,j,k} \rho_{g,i\pm 1/2,j} \cdot {}
\nonumber \\
& & \qquad x'_{i\pm 1/2,j} \left(\frac{df_{d,k}}{dx}\right)_{i\pm 1/2,j} \\ 
F_{{\rm diff},i,j\pm 1/2,k} & = & -D_{d,i\pm 1/2,j,k} \rho_{g,i,j\pm 1/2} \cdot {}
\nonumber \\
& & \qquad
\frac{\sqrt{1-\mu_{j\pm 1/2}^2}}{r_{i}} \left(\frac{df_{d,k}}{d\mu}\right)_{i,j\pm 1/2},
\end{eqnarray}
where $x' = dx/dr$ is the derivative of the mapping between radius and our radial variable $x$,
\begin{equation}
\left(\frac{df_{d,k}}{dx}\right)_{i\pm 1/2,j} = \frac{1}{\Delta x} \left(\frac{\rho_{d,i\pm 1,j,k}}{\rho_{g,i\pm 1,j}}-\frac{\rho_{d,ijk}}{\rho_{g,ij}}\right)
\end{equation}
and similarly for the $\mu$ derivatives. The subscripts indicate the cell face or centre at which all quantities are to be evaluated, and we note that $\rho_g$ and $D$ are known analytically at all positions.

We discretise the diffusion subsystem in time using a second-order accurate Crank-Nicolson scheme. Defining, $\Theta = 1/2$ as the time-centring parameter, for a time step $\Delta t$ we have
\begin{eqnarray}
\lefteqn{\frac{\rho_{d,ijk}^{(n+1)} - \rho_{d,ijk}^{(n)}}{\Delta t} = -\frac{1}{V_{ij}} \cdot {}}
\nonumber \\
& &
\left\{
\left(1-\Theta\right)
\left[
F_{{\rm diff},i+1/2,j,k}^{(n)} A_{i+1/2,j} - F_{{\rm diff},i-1/2,j,k}^{(n)} A_{i-1/2,j} + {}
\right.
\right.
\nonumber \\
& & \qquad\qquad
\left.
F_{{\rm diff},i,j+1/2,k}^{(n)} A_{i,j+1/2} - F_{{\rm diff},i,j-1/2,k}^{(n)} A_{i,j-1/2}
\right] + {}
\nonumber \\
& &
\;\;
\Theta \left[
F_{{\rm diff},i+1/2,j,k}^{(n+1)} A_{i+1/2,j} - F_{{\rm diff},i-1/2,j,k}^{(n+1)} A_{i-1/2,j} + {}
\right.
\nonumber \\
& & \qquad
\left.\left.
F_{{\rm diff},i,j+1/2,k}^{(n+1)} A_{i,j+1/2} - F_{{\rm diff},i,j-1/2,k}^{(n+1)} A_{i,j-1/2}
\right]\right\},
\label{eq:diff_discrete_b}
\end{eqnarray}
where $\rho_{d,ijk}^{(n)}$ and $\rho_{d,ijk}^{(n+1)}$ denote quantities evaluated at the previous and new times, respectively.

We can rearrange \autoref{eq:diff_discrete_b} to obtain a sparse linear system for each grain species $k$,
\begin{equation}
\matA_k \vecrho_{d,k}^{(n+1)} = \vecrho^{(n)}_k + \Delta t \dot{\vecrho}^{(n)}_{{\rm diff}, k}
\label{eq:diff_discrete_c}
\end{equation}
Here $\vecrho^{(n+1)}_{d,k}$ is vector with $N_r N_\theta$ elements ordered so that element $\ell$ contains the dust density in cell $(i,j) = (\ell \mod N_r, \lfloor\ell/N_r\rfloor)$, i.e.,
\begin{equation}
\rho^{(n+1)}_{\ell k} = \rho^{(n+1)}_{d,\ell\,\mathrm{mod}\,N_r, \lfloor \ell/N_r\rfloor,k},
\end{equation}
and similarly for $\vecrho^{(n)}_k$. The term $\dot{\vecrho}^{(n)}_{k,\rm diff}$ represents the rate of change in density due to diffusion evaluated at the old time, and is given by
\begin{eqnarray}
\lefteqn{
\left(\dot{\rho}_{\rm diff}\right)_{\ell k}^{(n)} = -\frac{1}{V_{ij}}\left(1-\Theta\right)\cdot {}
}
\nonumber \\
& & 
\left[
F_{{\rm diff},i+1/2,j,k}^{(n)} A_{i+1/2,j} - F_{{\rm diff},i-1/2,j,k}^{(n)} A_{i-1/2,j} + {}
\right.
\nonumber \\
& & \;\;
\left.
F_{{\rm diff},i,j+1/2,k}^{(n)} A_{i,j+1/2} - F_{{\rm diff},i,j-1/2,k}^{(n)} A_{i,j-1/2}
\right].
\end{eqnarray}

We solve \autoref{eq:diff_discrete_c} using a biconjugate gradient stabilised solver (BiCGSTAB) as implemented in the Eigen software package \citep{Guennebaud10a}.

\subsubsection{Advection}
\label{sssec:adv_subsystem}

The advection subsystem of \autoref{eq:advec_diff_discrete_a} is
\begin{eqnarray}
\lefteqn{
\frac{\partial}{\partial t} \rho_{d,ijk} = -V_{ij}^{-1} \cdot {}
}
\nonumber \\
& &
\left[
F_{{\rm adv},i+1/2,j,k} A_{i+1/2,j} - F_{{\rm adv},i-1/2,j,k} A_{i-1/2,j} + {}
\right.
\nonumber \\
& & \;\;
\left.
F_{{\rm adv},i,j+1/2,k} A_{i,j+1/2} - F_{{\rm adv},i,j-1/2,k} A_{i,j-1/2}
\right].
\label{eq:adv_discrete_a}
\end{eqnarray}
We solve this subsystem using the same TVD approach described in \aref{sssec:adv_subsystem_1d}. Given a starting state $\rho_{d,ijk}^{(n)}$, we advance the calculation from $t_{n}$ to $t_{n+1} = t_n + \Delta t$ via
\begin{eqnarray}
\rho_{d,ijk}^{(*)} & = & \rho_{d,ijk}^{(n)} + \Delta t \left(\dot{\rho}_{\rm adv}\right)_{d,ijk}^{(n)} \\
\rho_{d,ijk}^{(n+1)} & = & \frac{1}{2}\rho_{d,ijk}^{(n)} + \frac{1}{2}\rho_{d,ijk}^{(*)} + \frac{1}{2}\Delta t \left(\dot{\rho}_{\rm adv}\right)_{d,ijk}^{(*)},
\end{eqnarray}
where
\begin{eqnarray}
\lefteqn{\left(\dot{\rho}_{\rm adv}\right)_{ijk}^{(n)} =
-\frac{1}{V_{ij}}
\cdot {}
}
\nonumber \\
& &
\left[
F^{(n)}_{{\rm adv},i+1/2,j,k} A_{i+1/2,j} - F^{(n)}_{{\rm adv},i-1/2,j,k} A_{i-1/2,j} + {}
\right.
\nonumber \\
& & \;\;
\left.
F^{(n)}_{{\rm adv},i,j+1/2,k} A_{i,j+1/2} - F^{(n)}_{{\rm adv},i,j-1/2,k} A_{i,j-1/2}
\right]
\end{eqnarray}
is the rate of change in $\rho_{d,ijk}$ evaluated using the density field $\rho_{ijk}^{(n)}$, and similarly for $(\dot{\rho}_{\rm adv})_{ijk}^{(*)}$. We evaulate these terms in three steps.

First, for the initial density field $\rho_{ijk}^{(n)}$ or $\rho_{ijk}^{(*)}$ we compute a piecewise-parabolic method (PPM) reconstruction of the density field using the generalised PPM method of \citet{Skinner18a}, which extends the classical \citet{Colella84a} PPM method to curvilinear, non-uniform coordinates. Specifically, in the radial direction we approximate the density of grains in size bin $k$ as a function of position within cell $ij$ with a parabolic function
\begin{equation}
\rho_{d,k}(r) = c_{0,ijk} + c_{1,ijk} x(r) + c_{2,ijk} x(r)^2,
\end{equation}
where the reconstruction has the property that the average of $\rho_{d,k}(r)$ over the volume of cell $ij$ is $\rho_{d,ijk}$, and the function $\rho_{d,k}(r)$ is monotonic for $r \in (r_{i-1/2},r_{i+1/2})$. The reconstruction coefficients $c_0$, $c_1$, and $c_2$ are functions of $\rho_{d,i-1,jk}$, $\rho_{d,ijk}$, and $\rho_{d,i+1,jk}$, and chosen via the procedure described by \citet{Skinner18a}. The reconstruction is third-order accurate for smooth flows. We use the same procedure to obtain a PPM reconstruction of the density field in the $\mu$ direction.

Second, we calculate the velocities at cell faces. For each grain size bin we define $\beta_{k,0}$ by using $a_k$ in \autoref{eq:betazero}, and we discretize \autoref{eq:tau_geom} for the optical depth to cell edge $i+1/2$ along angle $j$ as
\begin{eqnarray}
\lefteqn{\tau_{i+1/2,j} = \sum_k \frac{3}{4a_k} \sum_{i'=0}^i }
\nonumber \\
& & \int_{r_{i'-1/2}}^{r_{i'+1/2}} \left(c_{0,i'jk} + c_{1,i'jk} x(r) + c_{2,i'jk} x(r)^2\right) \, dr.
\label{eq:tau_disc}
\end{eqnarray}
The integrals are trivial to evaluate analytically in each cell, since $x(r)$ is a known function. Given $\beta_{0,k}$ and $\tau_{i+1/2,j}$, along with the local Keplerian speed $\Omega_{\rm K}$, gas velocity $v_{g,\varpi}$, and stopping time $T_s$ (computed from our analytic background gas model), we can use \autoref{eq:vz} and \autoref{eq:vvarpi} to evaluate the $\varpi$ and $z$ velocities each face $i+1/2,j$, which we can trivially transform into the velocities $v_{d,r,i+1/2,jk}$ and $v_{d,\theta,i+1/2,jk}$ that we require. Our strategy for cell faces $i,j+1/2$ in the $\theta$ direction is similar, except that we cannot use the PPM reconstruction of the density field for rays along cell edges in the $\theta$ direction because it is not guaranteed to be continuous at cell edges. For this reason, we instead use an appropriately modiified version of \autoref{eq:tau_disc} to evaluate the optical depth to cell centres $ij$, and then take $\tau_{i,j+1/2} = (\tau_{ij} + \tau_{i,j+1})/2$. This then provides the velocities at the $i,j+1/2$ faces.

The third and final step is to calculate the rate of change terms. Consider the upper radial cell face, $i+1/2,j$, at which the velocity is $v_c = v_{r,i+1/2,j}$. After time $\Delta t$, the contact discontinuity between the two cells adjoining the face will be displaced from its initial location $r_c = r_{i+1/2,j}$ to a new location $r_c' = r_c-v_c\Delta t$, and thus, using our PPM reconstruction, the mass transported across the cell face during this time is
\begin{eqnarray}
\lefteqn{
\Delta M_{i+1/2,j,k} = 2\pi \Delta \mu 
\cdot
}
\nonumber \\
& &
\left\{
\begin{array}{ll}
\int_{r_c'}^{r_c} \left[c_{0,ijk} + c_{1,ijk}x(r) + c_{2,ijk} x(r)^2\right] r^2 \, dr, &
v_c > 0 \\
\int_{r_c}^{r_c'} \left[c_{0,i+1,j,k} + {}
\right.
\\
\qquad
\left.
c_{1,i+1,j,k}x(r) + c_{2,i+1,j,k} x(r)^2\right] r^2 \, dr, &
v_c < 0
\end{array}
\right.
\end{eqnarray}
The corresponding mass flux is $F_{{\rm adv},i+1/2,j,k} = \Delta M_{i+1/2,j,k}/A_{i+1/2,j}$. The expressions for the other three cell faces are analogous.

This completes a specification of the spatial discretization of \autoref{eq:advec_diff}; the full scheme we have described is second-order accurate in space.

\subsubsection{Floors}

In regions where the gas density is very low and the stopping time is large, grains can reach very high velocities, leading to very small time steps in \autoref{eq:timestep}. To avoid the computational cost of evolving cells with high speeds but that contain negligible mass, we apply a numerical floor. We first compute a floor density $\rho_{\rm floor}$ by requiring that the optical depth of a radial ray through the computational domain encountering dust in every size bin with density $\rho_{\rm floor}$ be $10^{-6}$, i.e.,
\begin{equation}
    \rho_{\rm floor} = 10^{-6} \left[\frac{3}{4} \left(r_{\rm max} - r_{\rm min}\right) \sum_k a_k^{-1} \right]^{-1}.
\end{equation}
For the choice of grain size and domain size used in all simulations presented in this paper, $\rho_{\rm floor} \approx 3.9 \times 10^{-26}$ g cm$^{-3}$.

We modify the update cycle by setting the advective flux $F_{\rm adv}$ (\autoref{sssec:adv_subsystem}) to zero across any cell interface for which the upwind density is below $\rho_{\rm floor}$. We similarly set the velocity across such interfaces to zero for the purposes of computing the time step (\autoref{eq:timestep}), and we do not require convergence in floored cells in our matrix solution in the diffusion subsystem (\autoref{sssec:diff_subsystem}).

\subsection{Boundary conditions}

Evaluation of the fluxes at the domain boundaries requires specification of boundary conditions. In the angular direction, we enforce zero advective and diffusive flux both across the midplane at $\mu=0$ and out of the top of the disc at $\mu = \mu_{\rm max}$. In terms of our spatial discretisation, this amounts to setting $v_{d,ijk,\theta} = 0$ and $D_{d,ijk} = 0$ for all $j=-1/2$ and $j=N_\theta-1/2$. In the radial direction, we use closed box boundary conditions for diffusion, and therefore set $D_{d,ijk} = 0$ for all $i=-1/2$ and $i=N_r-1/2$.

The advective radial flux requires a more sophisticated treatment. We  we wish to allow material to be advected inward across the inner radial boundary by drag, and to be pushed outward through the outer radial boundary by radiation. However, we do not wish new material to be able to enter the domain. We therefore adopt diode boundary conditions. At the inner boundary, we solve for the velocity across the innermost cell face $v_{d,-1/2,j,k,r}$ as described in the previous section, but if $v_{d,-1/2,j,k,r} > 0$ (i.e., if the velocity is radially outward, and thus into the domain), we set $v_{d,-1/2,j,k,r} = 0$ so that no mass enters the domain from smaller radii. We treat the outer boundary in the same way: we compute $v_{d,N_r-1/2,j,k,r}$, but if the resulting value is negative, indicating flow into the computational domain, we re-set the value to zero.


\bsp	
\label{lastpage}
\end{document}